\documentclass[prd,superscriptaddress,amsfonts,amssymb,amsmath,showpacs,twocolumn]{revtex4-2}
\usepackage{bm}
\usepackage{amsfonts}
\usepackage{latexsym}
\usepackage[latin1]{inputenc}
\usepackage{graphicx}
\usepackage{amsmath}
\usepackage{palatino}
\usepackage{ragged2e}
\usepackage{mathpazo}
\usepackage{textcomp}
\usepackage{float}
\usepackage{booktabs}
\usepackage{dcolumn}
\usepackage{multirow}
\usepackage{hyperref}
\hypersetup{colorlinks,citecolor=blue}
\usepackage{amsmath}
\usepackage{xcolor}
\usepackage{orcidlink}
\usepackage[caption=false]{subfig}
\usepackage{commath}
\def\jnl@style{\it}
\def\aaref@jnl#1{{\jnl@style#1}}

\def\aaref@jnl#1{{\jnl@style#1}}

\def\aj{\aaref@jnl{AJ}}                   % Astronomical Journal
\def\apj{\aaref@jnl{ApJ}}                 % Astrophysical Journal
\def\apjl{\aaref@jnl{ApJ}}                % Astrophysical Journal, Letters
\def\apjs{\aaref@jnl{ApJS}}               % Astrophysical Journal, Supplement
\def\apss{\aaref@jnl{Ap\&SS}}             % Astrophysics and Space Science
\def\aap{\aaref@jnl{A\&A}}                % Astronomy and Astrophysics
\def\aapr{\aaref@jnl{A\&A~Rev.}}          % Astronomy and Astrophysics Reviews
\def\aaps{\aaref@jnl{A\&AS}}              % Astronomy and Astrophysics, Supplement
\def\mnras{\aaref@jnl{Mon.~Not.~Roy.~Astron.~Soc.}}             % Monthly Notices of the RAS
\def\prd{\aaref@jnl{Phys.~Rev.~D}}        % Physical Review D
\def\prc{\aaref@jnl{Phys.~Rev.~C}}  % Physical Review C
\def\prl{\aaref@jnl{Phys.~Rev.~Lett.}}    % Physical Review Letters
\def\qjras{\aaref@jnl{QJRAS}}             % Quarterly Journal of the RAS
\def\skytel{\aaref@jnl{S\&T}}             % Sky and Telescope
\def\ssr{\aaref@jnl{Space~Sci.~Rev.}}     % Space Science Reviews
\def\zap{\aaref@jnl{ZAp}}                 % Zeitschrift fuer Astrophysik
\def\nat{\aaref@jnl{Nature}}              % Nature
\def\aplett{\aaref@jnl{Astrophys.~Lett.}} % Astrophysics Letters
\def\apspr{\aaref@jnl{Astrophys.~Space~Phys.~Res.}} % Astrophysics Space Physics Research
\def\physrep{\aaref@jnl{Phys.~Rep.}}      % Physics Reports
\def\physscr{\aaref@jnl{Phys.~Scr}}       % Physica Scripta
\def\commat{\aaref@jnl{Comm.~Math.~Phys.}}              % Communications in Mathematical Physics
\def\science{\aaref@jnl{Science}}               % Science
\def\cqg{\aaref@jnl{Classical Quant.~Grav.}}            % Classical and Quantum Gravity
\def\jpcs{\aaref@jnl{JPCS}}                                     % Journal of Physics Conference Series
\def\ijmpd{\aaref@jnl{Int.~J.~Mod.~Phys.~D}}                    % International Journal of Modern Physics D
\def\grg{\aaref@jnl{Gen.~Relat.~Gravit.}}               % General Relativity and Gravitation
\def\rpp{\aaref@jnl{Rep.~Prog.~Phys.}}          % Reports on Progress in Physics
\def\npa{\aaref@jnl{Nucl.~Phys.~A}}        % Nuclear Physics A
\def\lrr{\aaref@jnl{Living Rev.~Rel.}}                   % Living reviews in relativity
\def\jcap{\aaref@jnl{J.~Cosmology Astropart.~Phys.}}    % Journal of cosmology and astroparticle physics
\def\rmp{\aaref@jnl{Rev.~Mod.~Phys.}}   %Reviews of modern physics
\def\epjc{\aaref@jnl{Eur.~Phys.~J.~C}}

\allowdisplaybreaks[1]
\begin{document}

\color{black}       %% For one column

\title{Wormhole solutions in $f(R,L_m)$ gravity}

\author{Raja Solanki\orcidlink{0000-0001-8849-7688}}
\email{rajasolanki8268@gmail.com}
\affiliation{Department of Mathematics, Birla Institute of Technology and
Science-Pilani,\\ Hyderabad Campus, Hyderabad-500078, India.}
\author{Zinnat Hassan\orcidlink{0000-0002-6608-2075}}
\email{zinnathassan980@gmail.com}
\affiliation{Department of Mathematics, Birla Institute of Technology and
Science-Pilani,\\ Hyderabad Campus, Hyderabad-500078, India.}
\author{P.K. Sahoo\orcidlink{0000-0003-2130-8832}}
\email{pksahoo@hyderabad.bits-pilani.ac.in}
\affiliation{Department of Mathematics, Birla Institute of Technology and
Science-Pilani,\\ Hyderabad Campus, Hyderabad-500078, India.}
\affiliation{Faculty of Mathematics \& Computer Science, Transilvania University of Brasov, Eroilor 29, Brasov, Romania}

%%%%%%%%%%%%%%%%%%%%%%%%%%%%%%%%%%%%  DATE  %%%%%%%%%%%%%%%%%%%%%%%%%%%%%%%%%%%%
\date{\today}

\begin{abstract}

In this work, we intend to explore wormhole geometries in the framework of $f(R,L_m)$ gravity. We derive the field equations for the generic $f(R,L_m)$ function by assuming the static and spherically symmetric Morris-Thorne wormhole metric. Then we consider two non-linear $f(R,L_m)$ model, specifically, $f(R,L_m)=\frac{R}{2}+L_m^\alpha$ and $f(R,L_m)=\frac{R}{2}+(1+\lambda R)L_m$, where $\alpha$ and $\lambda$ are free model parameters. We obtain the wormhole solutions by assuming three cases, namely, a linear barotropic EoS, anisotropic EoS, and isotropic EoS corresponding to model I. We observe that for both barotropic and anisotropic cases, the corresponding wormhole solutions obey the flaring-out condition under asymptotic background, while for the isotropic case, the shape function does not follow the flatness condition. Also, we find that the null energy condition exhibits negative behavior in the vicinity of the throat. Further, we consider two different shape functions to investigate the behavior of model II. We find some constraints on the model parameter for which the violation of the null energy condition exhibits. Finally, we employ the volume integral quantifier to calculate the amount of exotic matter required near the wormhole throat for both models. We conclude that the modification of standard GR can efficiently minimize the use of exotic matter and provide stable traversable wormhole solutions.
\end{abstract}

\maketitle

\section{Introduction}\label{sec1}
In 1916, the idea of a wormhole was first suggested by L. Flamm \cite{Flamm}. Later on, Einstein and Rosen investigated the precise nature of a wormhole and made a hypothetical bridge by taking into account the idea of Flamm, known as the Einstein-Rosen bridge \cite{Einstein}. Wormholes are topologically framed hypothetical structures that supply a subway for distinct space-times apart from each other. A wormhole is supposed to be a tube-like structure that is asymptotically flat on both ends and connected by a throat. Wormholes are classified into two categories depending on the nature of the throat, namely static and non-static wormholes. A wormhole with a constant radius of the throat is referred to as a static wormhole, whereas a non-static wormhole represents a variable radius. Fuller and Wheeler demonstrated that one could not traverse through the Einstein-Rosen bridge, even a photon, since it would collapse instantly upon formation \cite{Fuller}. Further, Morris et al. \cite{Morris} proposed that exotic forms of matter threaded through a wormhole might hold it open; nevertheless, it remains to be seen whether such requirements are physically viable. Finally, in \cite{Thorne}, Morris and Thorne presented the static traversable wormholes that define interstellar travel and exact solutions of general relativity (GR). The authors show that a wormhole could be traversable if it exhibits exotic matter with a minimal surface area satisfying the flare-out condition. The matter content that violates null energy conditions (NEC) and defined by $T_{\mu\nu} k^\mu k^\nu \geq 0$ for any null vector $k^\mu$ (where $T_{\mu\nu}$ is the stress-energy tensor) is called the exotic matter, and hence it is an essential ingredient to construct a wormhole in GR; Indeed, in classical GR, wormhole solutions violate all energy conditions \cite{Visser}. Regardless, this type of hypothetical matter expresses unusuality in GR, whereas from the perspective of quantum gravity, it can be found as a natural consequence of fluctuations in the space-time topology \cite{Wheeler}. Thus, minimizing the violation of the energy conditions or reducing the quantity of exotic matter at the throat is essential. In scalar-tensor theories, wormhole solutions can be found for the scalar fields representing the function of phantom fluid \cite{Bronnikov1,Bronnikov2,Bronnikov3}. To bypass the undetected exotic matter, several modified gravity theories have appeared in the literature, such as $f(R)$ gravity \cite{Oliveira,Halilsoy,Azizi}, brane-world \cite{Kim,Camera,Riazi}, and curvature matter coupling \cite{Garcia1,Garcia2}. The effective stress-energy tensor involving the components of the geometry of space-time is now responsible for the violation of energy conditions.\\
An extension of the $f(R)$ modified gravity that incorporates an explicit coupling of the arbitrary function of the Ricci curvature $R$ with the matter Lagrangian term $L_m$ proposed in \cite{O.B.}. Harko and Lobo further generalized this case to arbitrary matter-geometry couplings \cite{THK}. The cosmological models with non-minimal curvature-matter couplings have significant astrophysical and cosmological applications \cite{THK-2,THK-3,THK-4,THK-5,V.F.-2}. Recently, Harko and Lobo proposed \cite{THK-6} $f(R,L_m)$ modified gravity that incorporates all curvature-matter coupling theories, where $f(R,L_m)$ is a generic function of the Ricci curvature $R$ and the Lagrangian term $L_m$. In the $f(R,L_m)$ modified gravity, the covariant divergence of the energy-momentum tensor does not vanishes, an extra force orthogonal to four velocities arises, and the motion of the test particle is non-geodesic. Further, the cosmological models with $f(R,L_m)$ gravity do not obey the equivalence principle, and that is constrained by the solar system experimental tests \cite{FR,JP}. Recently, several interesting cosmological and astrophysical works have been done in $f(R,L_m)$ gravity theory; for instance, one can check references \cite{GM,RV-1,RV-2,THK-7,THK-8}.\\
The equation of state (EoS) parameter plays a vital role in cosmology and astrophysics to describe the nature of the cosmic fluid. The fluid characterized by the linear barotropic EoS ($p=\omega \rho$) with positive energy density is found to be a viable candidate to describe cosmic evolution. In the context of GR, wormhole geometry with phantom energy background has been widely discussed \cite{Riazi,Gonzalez,Lobooo}. Further, in the context of modified gravity, such as $f(R)$, $f(R,T)$, and $f(Q)$ gravity, finding the exact solutions of the corresponding field equations is quite tricky compared to GR.\\
This paper studies wormhole geometry in the context of $f(R,L_m)$ gravity. We developed the corresponding field equations for the presence of the redshift function under the generic form of $f(R,L_m)$ gravity. Then we assumed two different $f(R,L_m)$ models and investigated wormhole solutions with various EoS relations and shape functions. This work is organized as follows: In sec \ref{sec2}, we present the fundamental formulations of $f(R,L_m)$ gravity theory. In sec \ref{sec3},  we derive the field equations for the generic $f(R,L_m)$ function corresponding to the static and spherically symmetric Morris-Thorne wormhole metric. Further in sec \ref{sec4}, we consider a specific $f(R,L_m)$ model, namely, $f(R,L_m)=\frac{R}{2}+L_m^\alpha$, where $\alpha$ is a free model parameter. Then, we study wormhole solutions for a linear barotropic EoS, an anisotropic EoS, and isotropic relation. In addition, we consider another non-linear $f(R,L_m)$ model, specifically, $f(R,L_m)=\frac{R}{2}+(1+\lambda R)L_m$ with two different shape functions $b(r)=r_0+\gamma ^2 r_0 \left(1-\frac{r_0}{r}\right)$ and $b(r)=r\,e^{r_0-r}$, and then we analyzed energy conditions. Further, in sec \ref{sec6}, we have discussed VIQ to check the amount of exotic matter required for a traversable wormhole. Finally, in the last section, we discuss our findings.

\section{ $f(R,L_m)$ Gravity Theory}\label{sec2}

\justify

The generic action governing the dynamics of the universe in $f(R,L_m)$ gravity read as

\begin{equation}\label{1a}
S= \int{f(R,L_m)\sqrt{-g}d^4x},
\end{equation}
where $R$ is the Ricci scalar corresponding to the metric $g_{\mu\nu}$ with determinant $g$ and $L_m$ represents the matter Lagrangian.\\
The Ricci scalar curvature term $R$ can be obtained by the contraction of the Ricci tensor $R_{\mu\nu}$ as
\begin{equation}\label{1b}
R= g^{\mu\nu} R_{\mu\nu},
\end{equation} 
where,
\begin{equation}\label{1c}
R_{\mu\nu}= \partial_\lambda \Gamma^\lambda_{\mu\nu} - \partial_\nu \Gamma^\lambda_{\lambda\mu} + \Gamma^\sigma_{\mu\nu} \Gamma^\lambda_{\sigma\lambda} - \Gamma^\lambda_{\nu\sigma} \Gamma^\sigma_{\mu\lambda},
\end{equation}
with $\Gamma^\alpha_{\beta\gamma}$ denoting the components of the Levi-Civita connection and can be calculated as
\begin{equation}\label{1d}
\Gamma^\alpha_{\beta\gamma}= \frac{1}{2} g^{\alpha\lambda} \left( \frac{\partial g_{\gamma\lambda}}{\partial x^\beta} + \frac{\partial g_{\lambda\beta}}{\partial x^\gamma} - \frac{\partial g_{\beta\gamma}}{\partial x^\lambda} \right).
\end{equation}

\justify The following field equation acquired by varying the generic action \eqref{1a} corresponding to the metric tensor $g_{\mu\nu}$,

\begin{multline}\label{1e}
f_R R_{\mu\nu} + (g_{\mu\nu} \square - \nabla_\mu \nabla_\nu)f_R - \frac{1}{2} (f-f_{L_m}L_m)g_{\mu\nu}\\
=\frac{1}{2} f_{L_m} T_{\mu\nu}.
\end{multline}
Here $f_R \equiv \frac{\partial f}{\partial R}$, $f_{L_m} \equiv \frac{\partial f}{\partial L_m}$, and $T_{\mu\nu}$ is the energy-momentum tensor for the cosmic fluid, given by 

\begin{equation}\label{1f}
T_{\mu\nu} = \frac{-2}{\sqrt{-g}} \frac{\delta(\sqrt{-g}L_m)}{\delta g^{\mu\nu}}.
\end{equation}
 Further, contraction of the field equation \eqref{1d} gives following relation between the energy-momentum scalar $T$, the Lagrangian term $L_m$, and the Ricci scalar $R$ as 
\begin{equation}\label{1g}
R f_R + 3\square f_R - 2(f-f_{L_m}L_m) = \frac{1}{2} f_{L_m} T,
\end{equation}
where $\square F = \frac{1}{\sqrt{-g}} \partial_\alpha (\sqrt{-g} g^{\alpha\beta} \partial_\beta F)$ for any scalar function $F$ .

\section{Wormhole Geometries in $f(R,L_m)$ Gravity}\label{sec3}

We consider the static and spherically symmetric Morris-Thorne wormhole metric \cite{Thorne,Visser}, given by
\begin{equation}\label{3a}
ds^2=-U(r)dt^2+V(r)dr^2+r^2d\Omega^2,
\end{equation}
Here, $d\Omega^2=d\theta^2+\text{sin}^2\theta d\phi^2$, $U(r)=e^{2\Phi(r)}$ and $V(r)=\left(1-\frac{b(r)}{r}\right)^{-1}$.
The functions $b(r)$ and $\Phi(r)$ of the radial coordinate encode the information about the shape of the wormhole and the gravitational redshift, and hence it is known as the shape function and the redshift function, respectively. To avoid the event horizon, $\Phi(r)$ must be finite everywhere. Moreover, to have traversable wormhole geometry, the shape function $b(r)$ should obey the following well-known constraints; specifically,  the flaring-out condition given by $(b-b'r)/b^2>0$ \cite{Thorne},  throat condition given by $b(r_0)=r_0$ ($r_0$ is the throat radius) with $b^{\,\prime}(r_0)<1$, and the asymptotic flatness condition, that is given by  $\frac{b(r)}{r}\rightarrow 0$ as $r\rightarrow \infty$. Also, another significant criterion is the proper radial distance $l(r)$, represented as
\begin{equation}
l(r)=\pm \int_{r_0}^{r}\frac{dr}{\sqrt{1-\frac{b(r)}{r}}},
\end{equation}
is needed to be finite everywhere. Here, the $\pm$ symbols indicate the upper and lower portions of the wormhole, which are linked by the throat. Also, the proper distance decreases from the upper universe $l=+\infty$ to
the throat and then from  $l=0$ to $-\infty$ in the lower universe. Moreover, $l$ should be greater than or equal to the coordinate distance $\mid l(r)\mid \geq r-r_0$. The embedding surface of the wormhole can be obeyed by defining the embedding surface $z(r)$ at a fixed $\theta=\pi/2$ and time $t=\text{constant}$. Hence, equation \eqref{3a} reduces to
\begin{equation}
\label{6a}
ds^2=\left(1-\frac{b(r)}{r}\right)^{-1}dr^2+r^2 d\phi^2.
\end{equation}
The above metric can be embedded into three-dimensional Euclidean space with cylindrical coordinates $r,\,\ \phi$ and $z$ as
\begin{equation}
\label{6b}
ds^2=dz^2+dr^2+r^2 d\phi^2.
\end{equation}
Now, on comparing equations \eqref{6a} and \eqref{6b}, we obtained the following slope equation so that by integrating it, one can find the embedding surface $z(r)$,
\begin{equation}
\label{6c}
\frac{dz}{dr}=\pm \sqrt{\frac{r}{r-b(r)}-1}.
\end{equation}
%In GR, having a traversable wormhole geometry leads to the presence of exotic matter at the wormhole's throat.\\
\\
Now corresponding to the metric \eqref{3a}, we obtained the non-vanishing components of the Ricci tensor as
\begin{multline}
R_{00}=e^{2\Phi} \left[ \left( 1-\frac{b}{r} \right)  \left\lbrace \Phi''+\Phi'^2+\frac{2\Phi'}{r} \right\rbrace \right.\\\left.
- \frac{(rb'-b)}{2r^2}\Phi'  \right],
\end{multline}
\begin{equation}
R_{11}=  - \Phi''-\Phi'^2  + \frac{(rb'-b)}{2r(r-b)} \left( \Phi' +\frac{2}{r} \right),
\end{equation}
\begin{equation}
R_{22}= (b-r)\Phi' + \frac{b'}{2} + \frac{b}{2r},
\end{equation}
\begin{equation}
R_{33}= \sin^2\theta \left\lbrace (b-r)\Phi' + \frac{b'}{2} + \frac{b}{2r} \right\rbrace ,
\end{equation}
Then by using equation \eqref{1b}, we obtained the Ricci curvature scalar $R$ for the spherically symmetric configuration \eqref{3a} as
\begin{multline}\label{3c}
R=\frac{2b'}{r^2} - 2\left( 1-\frac{b}{r} \right) \left\lbrace \Phi''+\Phi'^2+\frac{\Phi'}{r} \right\rbrace \\
+ \frac{\Phi'}{r^2} \left( rb'+b-2r \right).
\end{multline}
The stress-energy tensor corresponding to anisotropic fluid is defined by
\begin{equation}\label{3b}
T_{\mu\,\nu}=\left(\rho+P_t\right)u_{\mu}\,u_{\nu}+P_t\,\delta_{\mu\,\nu}+\left(P_r-P_t\right)v_{\mu}\,v_{\nu},
\end{equation}
where, $u_{\mu}$ and $v_{\mu}$ represent the four-velocity vector and the unitary space-like vector, respectively and they satisfy the conditions $u_{\mu}u^{\nu}=-v_{\mu}v^{\nu}=-1$. Here, $\rho$ is the energy density, $P_r$, and $P_t$ are the radial and tangential pressure, depending on the radial coordinate $r$ only. In the case of a stress-energy tensor of the form \eqref{3b}, the NEC can be defined as $\rho+p_i\geq 0$ (where $i$ is either $r$ or $t$).\\
Now, inserting the metric \eqref{3a} and the anisotropic
fluid \eqref{3b}, into the equations of motion \eqref{1e},
we obtained the following field equations, which read as
\begin{multline}\label{3d}
\left( 1-\frac{b}{r} \right) \left[ \left\lbrace  \Phi''+\Phi'^2  + \frac{2\Phi'}{r} - \frac{(rb'-b)}{2r(r-b)}\Phi' \right\rbrace F -\left\lbrace \Phi' \right.\right.\\\left.\left.
+\frac{2}{r}- \frac{(rb'-b)}{2r(r-b)}  \right\rbrace F' - F'' \right] + \frac{1}{2} \left( f-L_m f_{L_m} \right) = \frac{1}{2} f_{L_m} \rho,
\end{multline}
\begin{multline}\label{3e}
\left( 1-\frac{b}{r} \right) \left[ \left\lbrace  - \Phi''-\Phi'^2  + \frac{(rb'-b)}{2r(r-b)} \left( \Phi' +\frac{2}{r} \right) \right\rbrace F  \right.\\\left.
+ \left\lbrace \Phi'+\frac{2}{r}- \frac{(rb'-b)}{2r(r-b)}  \right\rbrace F' \right] - \frac{1}{2} \left( f-L_m f_{L_m} \right) = \frac{1}{2} f_{L_m} p_r,
\end{multline}
\begin{multline}\label{3f}
\left( 1-\frac{b}{r} \right) \left[ \left\lbrace - \frac{\Phi'}{r}  + \frac{(rb'+b)}{2r^2(r-b)} \right\rbrace F+ \left\lbrace \Phi'+\frac{2}{r} \right.\right.\\\left.\left.
- \frac{(rb'-b)}{2r(r-b)}  \right\rbrace F' + F'' \right] - \frac{1}{2} \left( f-L_m f_{L_m} \right) = \frac{1}{2} f_{L_m} p_t.
\end{multline}
where $F=\frac{\partial f}{\partial R}$.
Finally, we end up with a set of three equations (\ref{3d}-\ref{3f}) involving six unknown quantities. Thus the above system of equations is under-determined, and we need some extra constraints to construct the wormhole solutions.

\section{Wormhole Solutions for some specific Non-linear $f(R,L_m)$ Models}\label{sec4}
 The $f(R,L_m)$ function that we have considered in our study is motivated by the generic $f(R,L_m)$ function; specifically,  $f(R,L_m) = f_1(R)+f_2(R) G(L_m)$ that represents arbitrary curvature-matter coupling \cite{LB}. The models with minimal and non-minimal couplings have received great attention from cosmologists in the recent past in the context of different modified gravities such as $f(R,\mathcal{T})$, $f(T,\mathcal{T})$, and $f(Q,\mathcal{T})$ gravity. The above generic $f(R,L_m)$ functions are attractive since they consist of minimal and non-minimal coupling cases. We will explore wormhole geometry by incorporating both minimal and non-minimal coupling cases. 
 
\subsection{Model I}\label{subsec1}

We presume the following minimal $f(R,L_m)$ function to obtain the wormhole solutions\cite{LB,Jay-2}. The type of minimal coupling case we considered here is motivated by an interesting work of Bose et al. \cite{Sube} in the context of $f(R,\mathcal{T})$ gravity.
\begin{equation}\label{3g} 
f(R,L_m)=\frac{R}{2}+L_m^\alpha.
\end{equation}
\justify where $\alpha$ is a free model parameter. In particular, for the case $\alpha=1$, we retrieve the usual wormhole geometry of GR.\\
Therefore, we obtain the following field equations corresponding to this specific $f(R,L_m)$ function,  
\begin{equation}\label{3h}
\frac{b'}{r^2} = (2\alpha-1) \rho^\alpha,
\end{equation}
\begin{equation}\label{3i}
-\frac{b}{r^3} + 2 \left( 1-\frac{b}{r} \right) \frac{\Phi'}{r} = \rho^{\alpha-1} \left\lbrace (1-\alpha)\rho + \alpha p_r  \right\rbrace ,
\end{equation}
\begin{multline}\label{3j}
\left( 1-\frac{b}{r} \right) \left[ \Phi''+\Phi'^2+ \left\lbrace \frac{1}{r} - \frac{(rb'-b)}{2r(r-b)} \right\rbrace \Phi' \right.\\\left.
- \frac{(rb'-b)}{2r^2(r-b)} \right] = \rho^{\alpha-1} \left\lbrace (1-\alpha)\rho + \alpha p_t  \right\rbrace .
\end{multline}
In this work, we incorporate the constant redshift function $\Phi(r)=\text{constant}$ to obtain the wormhole solution. Now in order to find the analytic solutions of the obtained field equations, we need only one extra ansatz. Therefore, we consider the following cases:

\subsubsection{Linear barotropic EoS}\label{subsub11}
\begin{figure}[H]
    \centering
    \includegraphics[scale=0.6]{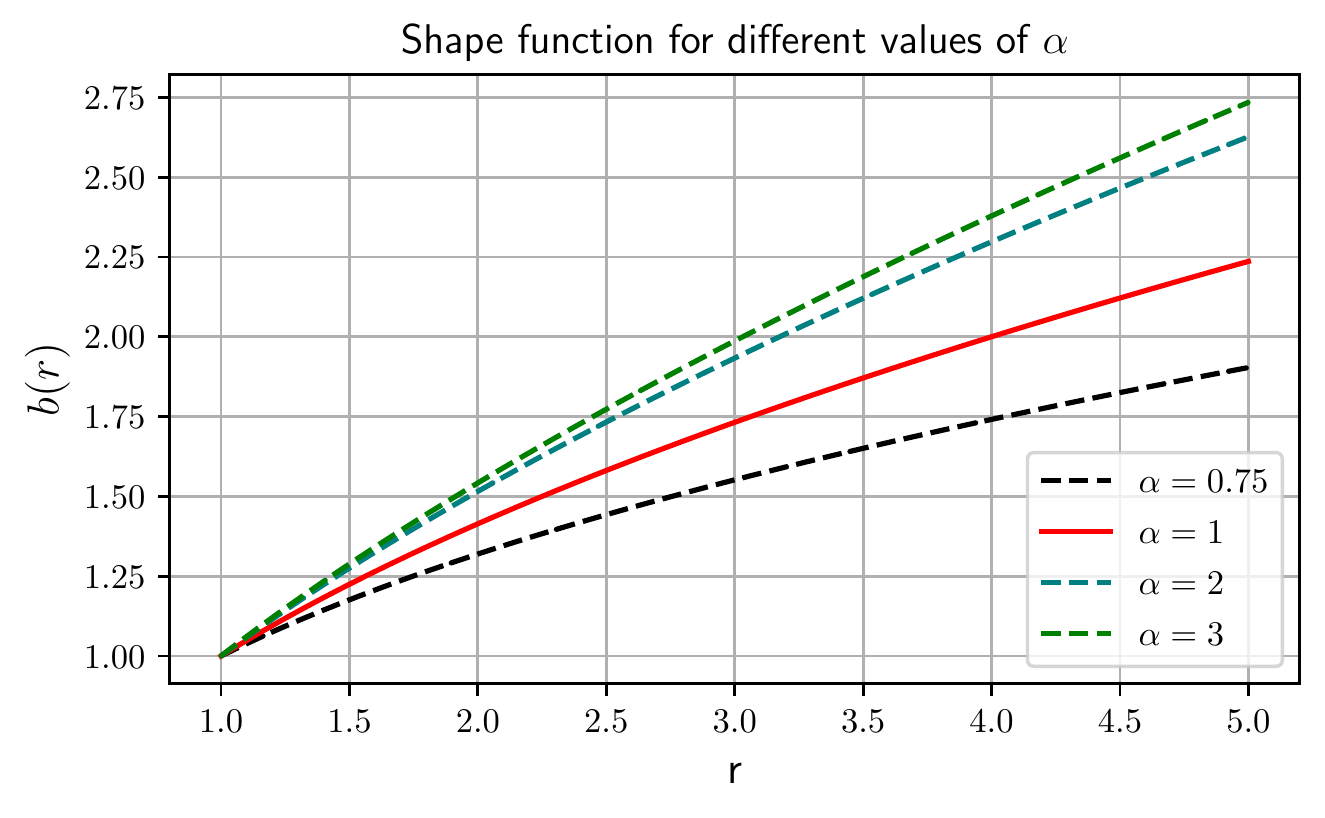}
    \caption{Shape function with $\omega=-2$ and $r_0=1$ ($p_r=\omega \rho$).}
    \label{fig:1}
\end{figure}

\begin{figure}[H]
    \centering
    \includegraphics[scale=0.6]{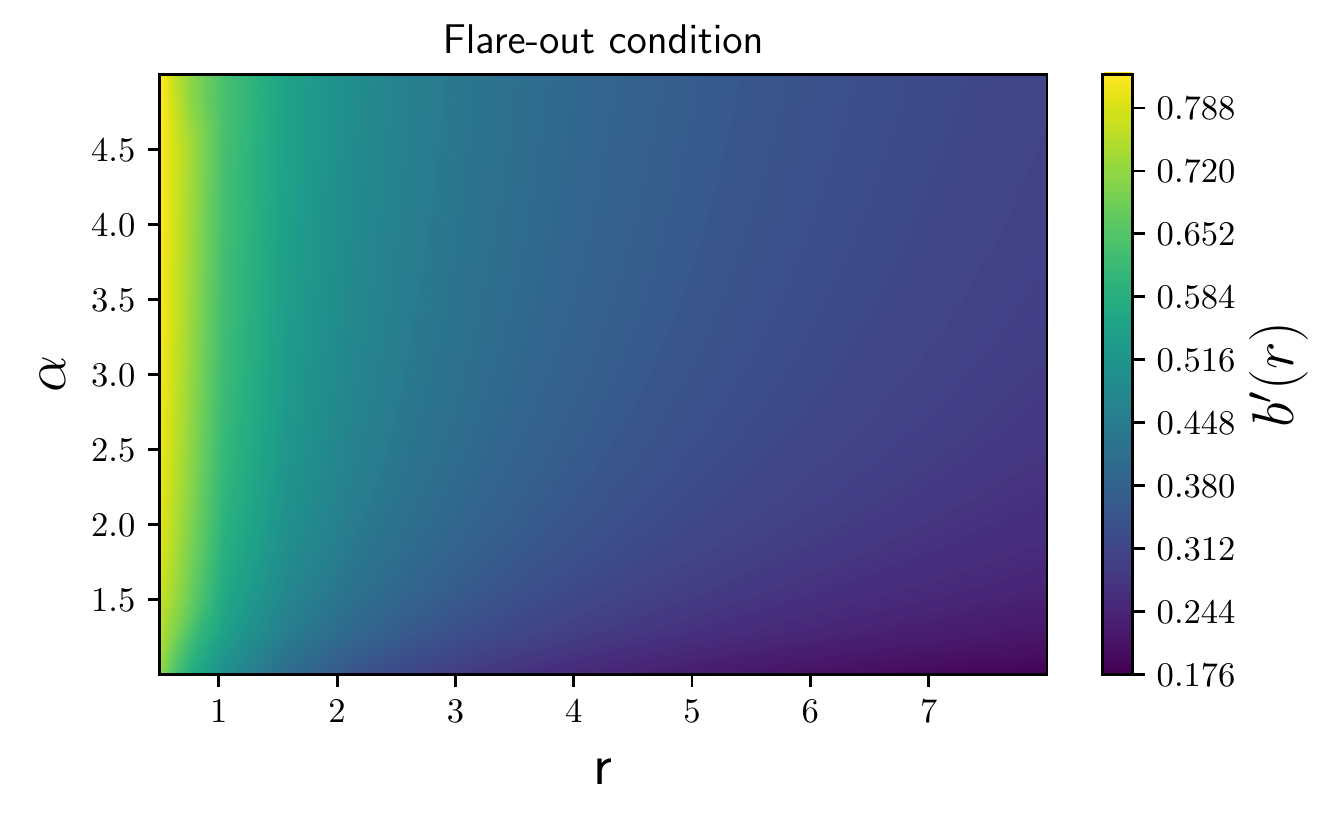}
    \caption{Flare-out condition with $\omega=-2$ and $r_0=1$ ($p_r=\omega \rho$).}
    \label{fig:2}
\end{figure}

\begin{figure}[h]
\centering
\includegraphics[width=8cm, height=6cm]{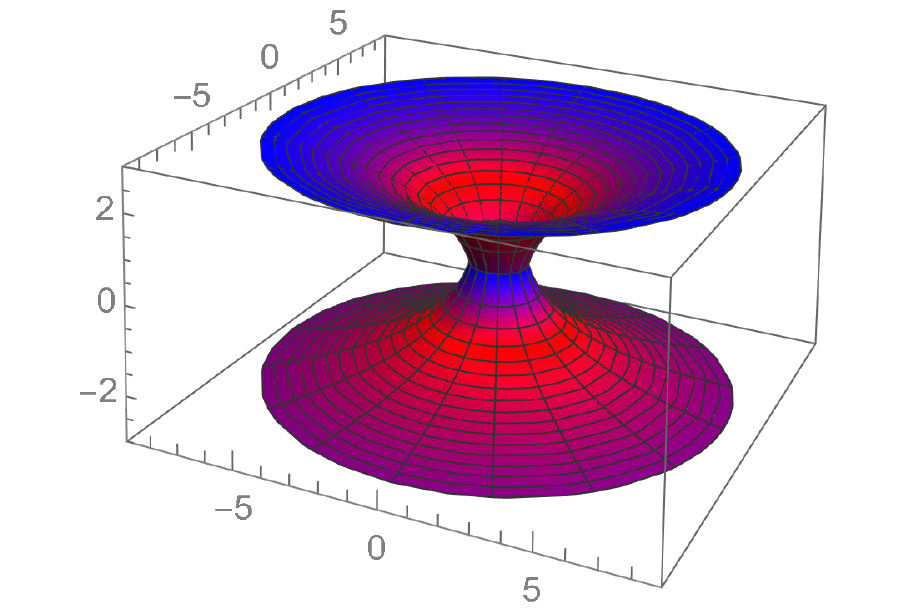}
\caption{Embedding diagram for the linear barotropic EoS model corresponding to parameter values $\omega=-2$ and $\alpha=0.4$.}
\label{fig:emb1}
\end{figure}

\justify In this particular section, we are going to construct wormhole solutions by considering the following linear barotropic EoS \cite{Lobo1}.  
\begin{equation}\label{4a}
p_r=\omega \rho
\end{equation}
where $\omega$ is the EoS parameter. In the context of late-time cosmic acceleration, the $\Lambda$CDM description of the dark energy is characterized by the EoS parameter $\omega=-1$ and is the most successful theory so far. Another widely discussed time-dependent dark energy
model is the quintessence model characterized by the EoS parameter $\omega>-1$, whereas the least theoretically understood dark energy is the phantom energy characterized by $\omega<-1$.
In Ref. \cite{Lobo2}, the authors investigated the asymptotically flat phantom wormhole solutions. Also, it is mentioned in \cite{Mandal,Jasim} that finding wormhole solutions with linear barotropic EoS in the framework of a non-linear model is quite difficult in both teleparallel and symmetric teleparallel gravity. However, we obtain the exact wormhole solution in the framework of our non-linear $f(R,L_m)$ model with linear barotropic EoS. \\
On solving the equations \eqref{3h} and \eqref{3i} by incorporating the relation \eqref{4a}, we obtained the following first order differential equation
\begin{equation}\label{4b}
b'+ \frac{(2\alpha-1)}{(1-\alpha+\alpha \omega)} \frac{b}{r}=0.
\end{equation}
Now on integrating the above equation by using the throat condition $b(r_0)=r_0$, we obtained the following expression for the shape function $b(r)$ in terms of radial coordinate r
\begin{equation}\label{4c}
b(r)=r_0  \left( \frac{r_0}{r} \right)^{\frac{(2\alpha-1)}{(1-\alpha+\alpha \omega)}}.
\end{equation}
The behavior of the shape function given in equation \eqref{4c} and the flaring-out condition are presented in Figs. \ref{fig:1} and \ref{fig:2}.
From our investigation, we found that $\frac{b(r)}{r}\rightarrow 0$ as $r\rightarrow \infty$ will satisfy only if (i) $\omega<-1$ with $\alpha>-\frac{1}{\omega-1}$, and (ii) $-1<\omega<1$ with $0<\alpha<-\frac{1}{\omega-1}$. As we are interested in exploring wormhole geometry under the phantom scenario, we consider the range $\omega<-1$ with $\alpha>-\frac{1}{\omega-1}$ in this analysis. By setting the appropriate values from the above region, we have plotted the shape functions in Figs. \ref{fig:1} and \ref{fig:2}. We found that the flaring-out condition is satisfied with asymptotic flatness.\\
Now, by using the equation \eqref{4c} in equations \eqref{3h}-\eqref{3j}, one can acquire the expressions for the energy density, radial pressure, and the tangential pressure as
\begin{equation}\label{4d}
\rho(r)=\left(-\frac{r_0 \left(\frac{r_0}{r}\right)^{\frac{2 \alpha -1}{\alpha  (\omega -1)+1}}}{r^3 (\alpha  (\omega -1)+1)}\right)^{1/\alpha },
\end{equation}
\begin{equation}\label{4e}
p_r(r)=\omega  \left(-\frac{r_0 \left(\frac{r_0}{r}\right)^{\frac{2 \alpha -1}{\alpha  (\omega -1)+1}}}{r^3 (\alpha  (\omega -1)+1)}\right)^{1/\alpha },
\end{equation}
\begin{equation}\label{4f}
p_t(r)=-\frac{(\alpha  (\omega -1)+2) \left(-\frac{r_0 \left(\frac{r_0}{r}\right)^{\frac{2 \alpha -1}{\alpha  (\omega -1)+1}}}{r^3 (\alpha  (\omega -1)+1)}\right)^{1/\alpha }}{2 \alpha }.
\end{equation}
Now we analyze the behavior of Null energy condition (NEC) in the vicinity of the throat.  
The NEC corresponding to our linear barotropic EoS model read as
\begin{equation}\label{4g}
\rho+p_r=(\omega +1) \left(-\frac{r_0 \left(\frac{r_0}{r}\right)^{\frac{2 \alpha -1}{\alpha  (\omega -1)+1}}}{r^3 (\alpha  (\omega -1)+1)}\right)^{1/\alpha },
\end{equation}
\begin{equation}\label{4h}
\rho+p_t=-\frac{(\alpha  (\omega -3)+2) \left(-\frac{r_0 \left(\frac{r_0}{r}\right)^{\frac{2 \alpha -1}{\alpha  (\omega -1)+1}}}{r^3 (\alpha  (\omega -1)+1)}\right)^{1/\alpha }}{2 \alpha }.
\end{equation}
At the throat, the above expressions reduce to the following,
\begin{equation}\label{4i}
\rho+p_r\mid_{r=r_0}=(\omega +1) \left(-\frac{1}{r_0^2 (\alpha  (\omega -1)+1)}\right)^{1/\alpha },
\end{equation}
\begin{equation}
\rho+p_t\mid_{r=r_0}=-\frac{(\alpha  (\omega -3)+2) \left(-\frac{1}{r_0^2 (\alpha  (\omega -1)+1)}\right)^{1/\alpha }}{2 \alpha }.
\end{equation}
\begin{figure}[h]
    \centering
    \includegraphics[scale=0.6]{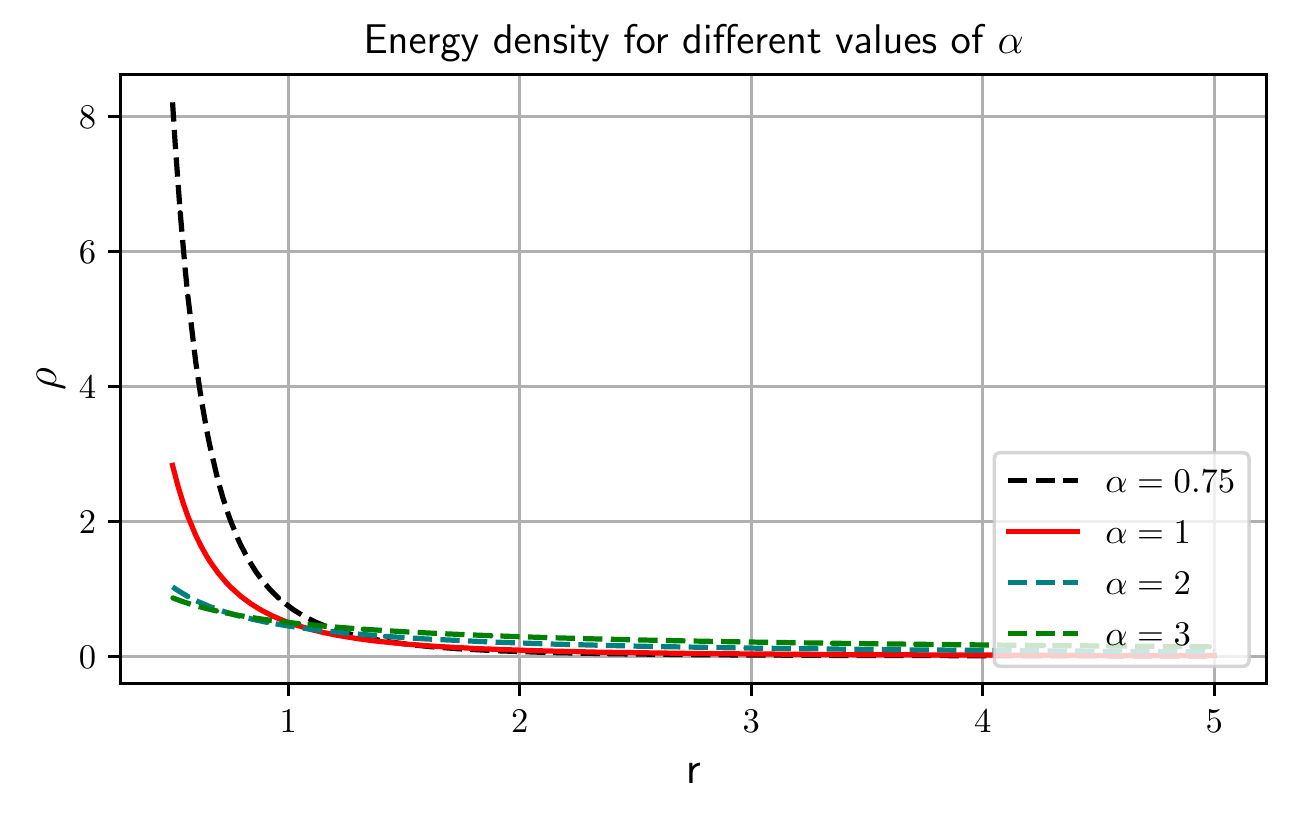}
    \caption{Energy density with $\omega=-2$ and $r_0=1$ ($p_r=\omega \rho$).}
    \label{fig:3}
\end{figure}
\begin{figure}[h]
    \centering
    \includegraphics[scale=0.6]{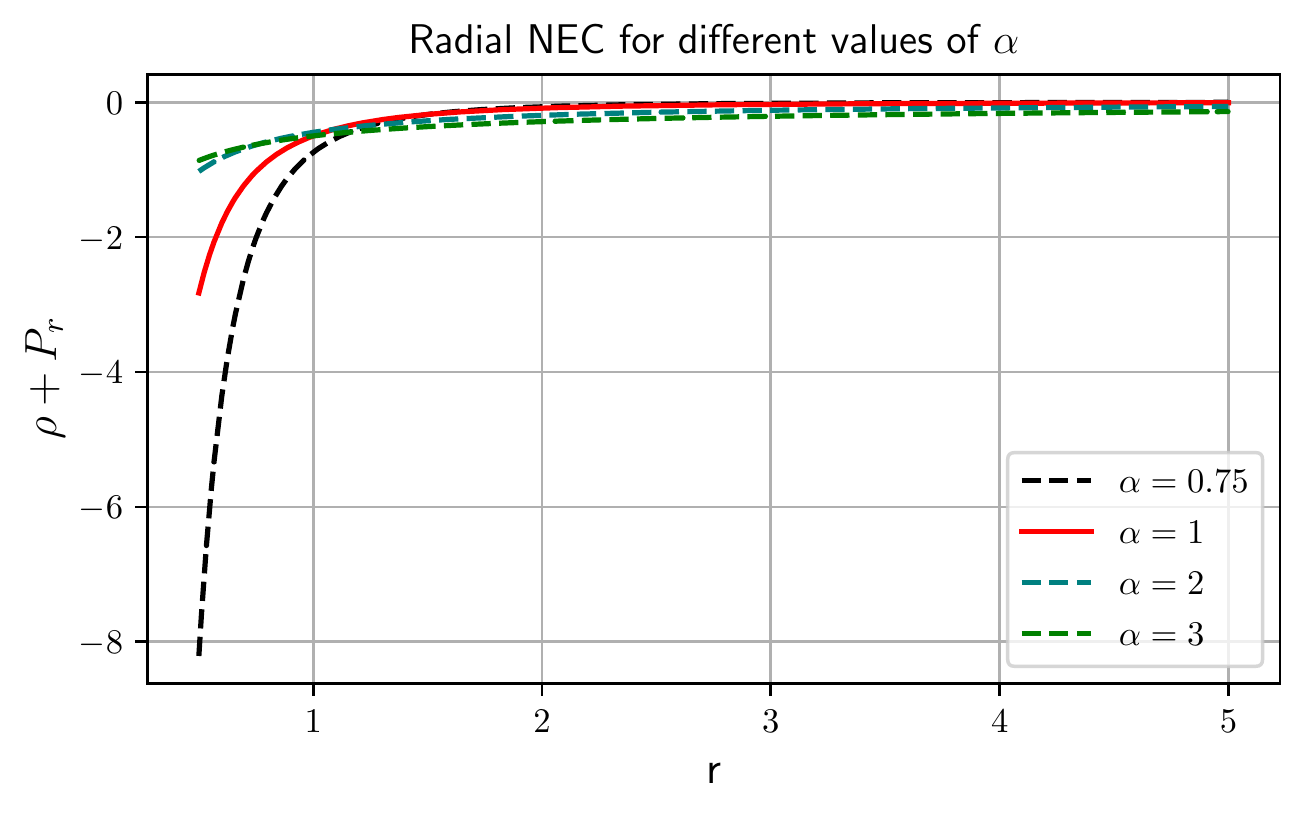}
    \caption{Radial NEC with $\omega=-2$ and $r_0=1$ ($p_r=\omega \rho$).}
    \label{fig:4}
\end{figure}
We have presented the behavior of the energy density and radial NEC in Fig. \ref{fig:3} and \ref{fig:4}, respectively. We found that the energy density shows positive behavior, whereas the radial NEC exhibits negative behavior. Also, one can observe that the right-hand side of the \eqref{4i} is a negative quantity corresponding to positive values of $\alpha$ with $\omega<-1$, which confirms the violation of NEC at the wormhole throat. Moreover, we have obtained the embedding surface for this case by using the Eq. \eqref{6c}, which is shown in Fig.\ref{fig:emb1}.
%\begin{figure}[H]
%    \centering
%    \includegraphics[scale=0.6]{Model-5.pdf}
%    \caption{SEC with $\omega=-2$ and $r_0=1$ ($p_r=\omega \rho$).}
%    \label{fig:5}
%\end{figure}

\subsubsection{Anisotropic EoS}\label{subsub12}
\begin{figure}[H]
    \centering
    \includegraphics[scale=0.6]{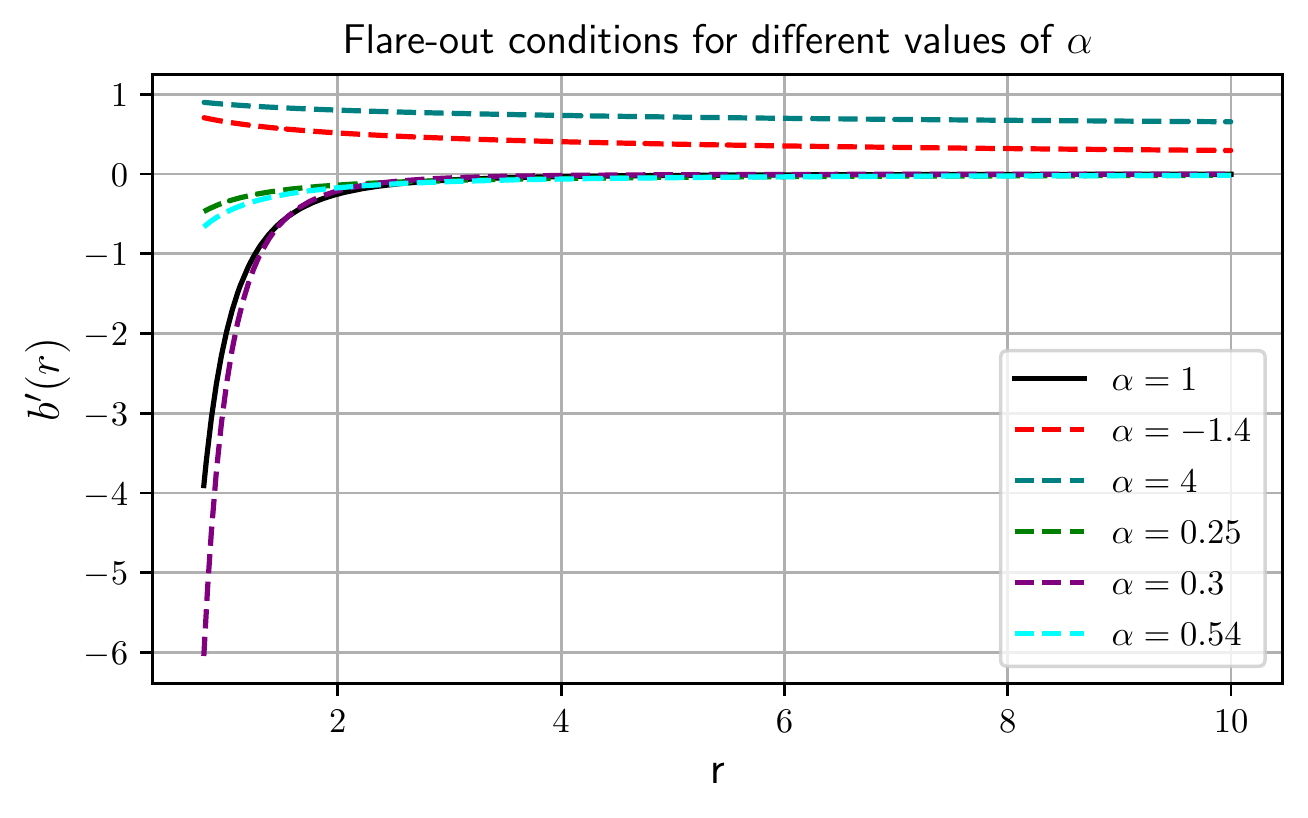}
    \caption{Flare-out condition with $r_0=1$ ($p_t=n p_r$). Here, we plotted for the parameter values $(\alpha,n)$: $(1,-1.5)$, $(-1.4,-1)$, $(4,-0.7)$, $(0.25,-0.1)$, $(0.3,0.5)$, $(0.54,2)$.}
    \label{fig:5}
\end{figure}

\begin{figure}[h]
    \centering
    \includegraphics[scale=0.59]{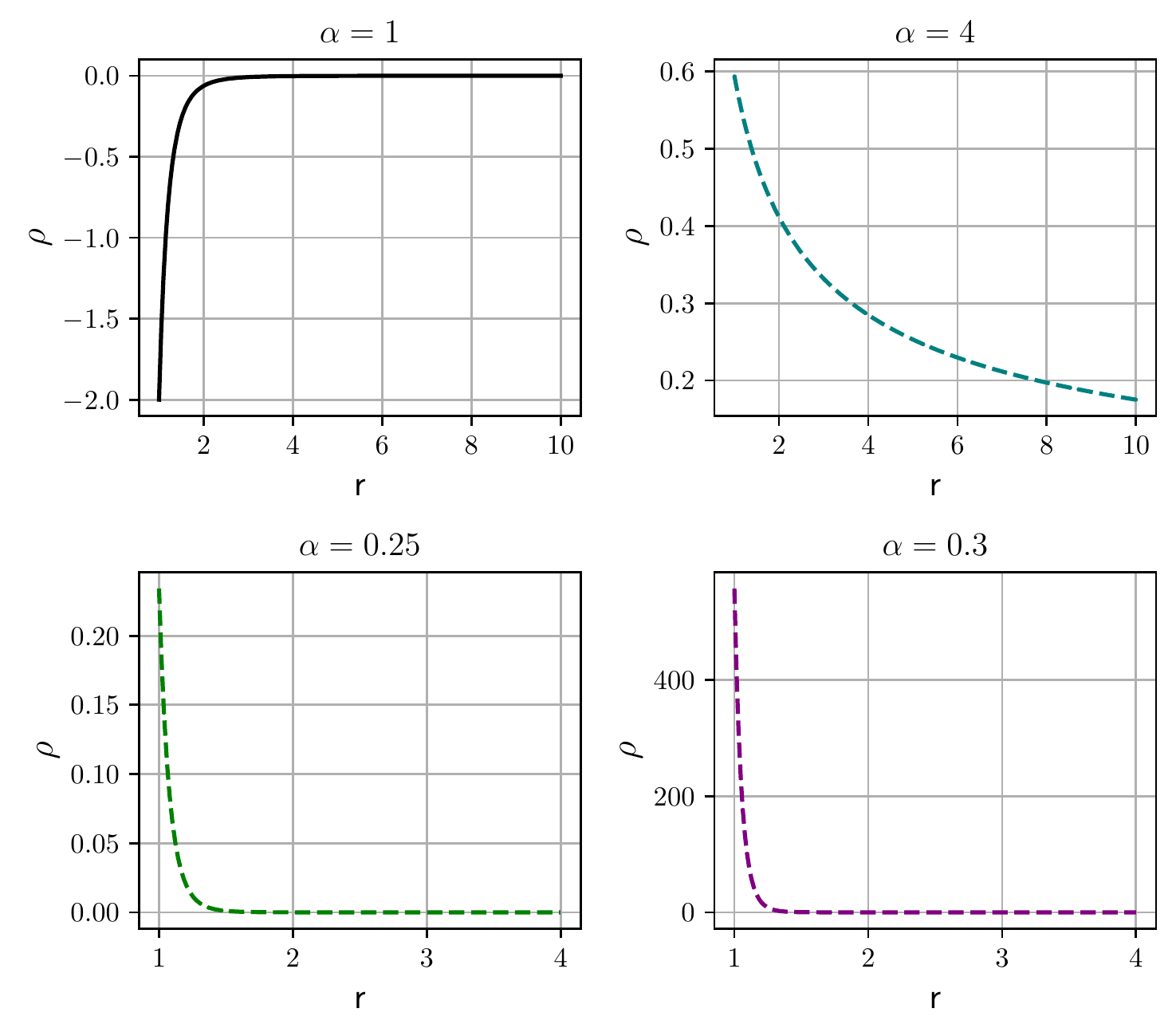}
    \caption{Energy density with $r_0=1$ ($p_t=n p_r$). Here, we plotted for the parameter values $(\alpha,n)$: $(1,-1.5)$, $(4,-0.7)$, $(0.25,-0.1)$, $(0.3,0.5)$.}
    \label{fig:6}
\end{figure}

\begin{figure}[h]
   \centering
    \includegraphics[scale=0.59]{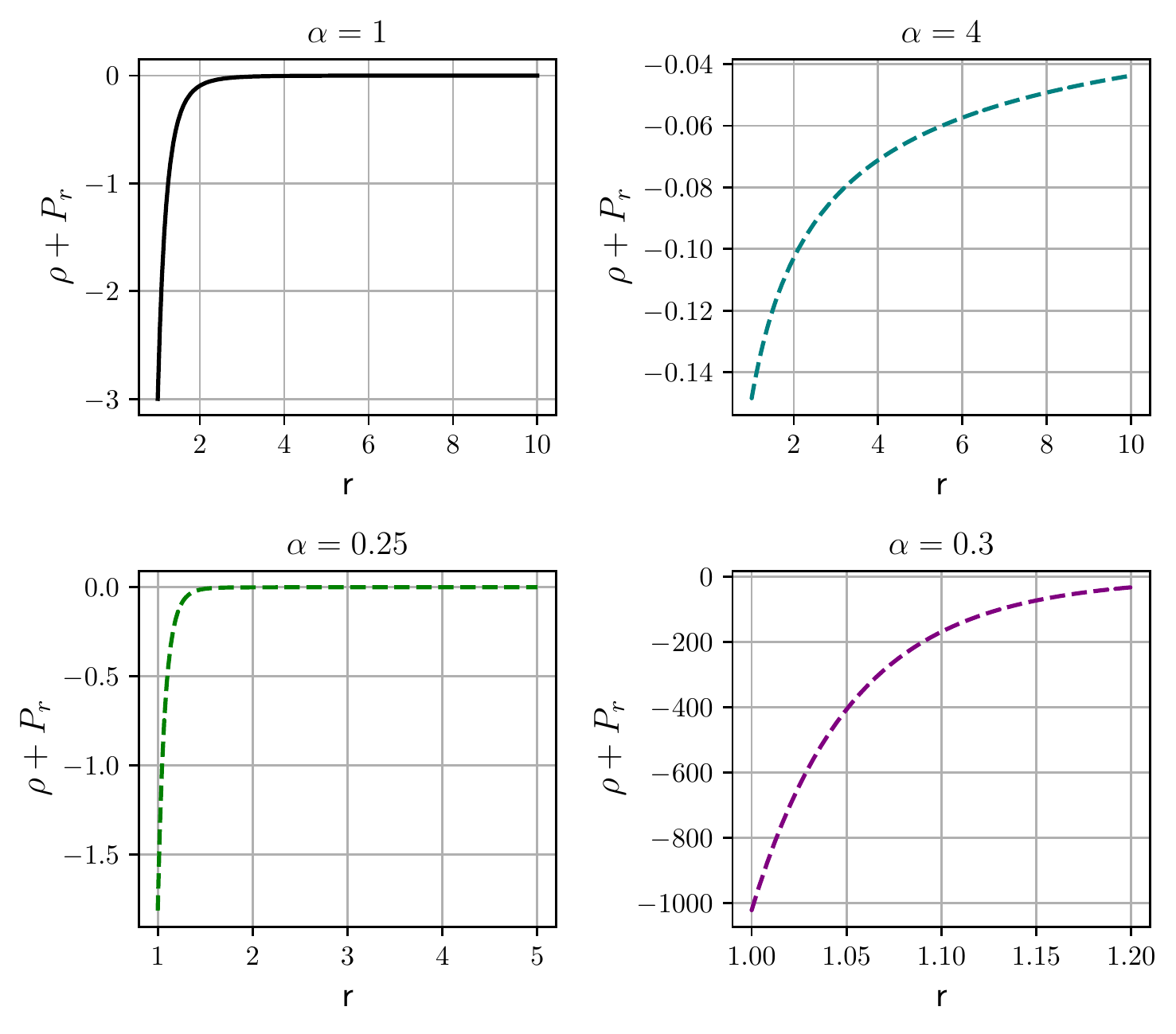}
    \caption{Radial NEC with $r_0=1$ ($p_t=n p_r$). Here, we plotted for the parameter values $(\alpha,n)$: $(1,-1.5)$, $(4,-0.7)$, $(0.25,-0.1)$, $(0.3,0.5)$.}
   \label{fig:7}
\end{figure}

\justify 
In this subsection, we consider the following anisotropic EoS to construct wormhole solutions \cite{Moraes1}.  
\begin{equation}\label{5a}
p_t=n p_r,
\end{equation}
where $n \neq 1$. On solving the equations \eqref{3h}-\eqref{3j} by incorporating the relation \eqref{5a}, we obtained the following first-order differential equation
\begin{equation}\label{5b}
b'- \frac{(2n+1)(2\alpha-1)}{\left\lbrace 2n(\alpha-1)+1 \right\rbrace} \frac{b}{r}=0.
\end{equation}
Now on integrating the above equation by using the throat condition $b(r_0)=r_0$, we obtained the following expression for the shape function $b(r)$ 
\begin{equation}\label{5c}
b(r)=r_0  \left( \frac{r}{r_0} \right)^{\frac{(2\alpha-1)(2n+1)}{[2n(\alpha-1)+1]}}.
\end{equation}
From our investigation, we found some constraints on parameters $\alpha$ and $n$ to satisfy the asymptotic flatness condition corresponding to the shape function obtained in the equation \eqref{5c}. We listed out the specific constraints in Table \ref{Table-1}.
\begin{widetext}
\begin{table}[H]
\begin{center}
\begin{tabular}{|c|c|c|c|c|c|c|c|}
\hline
$n$ & $n<-1$ & $n=-1$  & $-1< n \leq -\frac{1}{2} $ & $-\frac{1}{2} < n < 0$ & $0< n < 1$ & $n>1$\\
\hline 
$\alpha$ & $\alpha \in \left( \frac{1}{n+1} , \frac{2n-1}{2n} \right)$ & $\alpha < \frac{3}{2}$ & $\alpha < \frac{2n-1}{2n}$ or $\alpha > \frac{1}{n+1}$ & $\alpha > \frac{2n-1}{2n}$ or $\alpha < \frac{1}{n+1}$ & $\alpha \in \left( \frac{2n-1}{2n} , \frac{1}{n+1} \right)$ & $\alpha \in \left( \frac{1}{n+1} , \frac{2n-1}{2n} \right)$\\
\hline
\end{tabular}
\caption{Table shows the acceptable ranges to satisfy the asymptotic flatness condition.}
\label{Table-1}
\end{center}
\end{table}
\end{widetext}
Now we study the flaring-out condition for the shape function \eqref{5c}. By choosing some appropriate values from Table \ref{Table-1}, we have shown the behavior of the flaring-out condition in Fig. \ref{fig:5}. One can notice that the flaring-out condition is satisfied at the wormhole throat.
Now, by using the equation \eqref{5c} in equations \eqref{3h}-\eqref{3j}, one can acquire the expressions for the energy density, radial pressure, and the tangential pressure as
\begin{equation}\label{5d}
\rho(r)=\left(\frac{(2 n+1) r_0 \left(\frac{r}{r_0}\right)^{\frac{(2 \alpha -1) (2 n+1)}{2 (\alpha -1) n+1}}}{r^3 (2 (\alpha -1) n+1)}\right)^{1/\alpha },
\end{equation}
\begin{equation}\label{5e}
p_r(r)=\frac{(\alpha -2) \left(\frac{(2 n+1) r_0 \left(\frac{r}{r_0}\right)^{\frac{(2 \alpha -1) (2 n+1)}{2 (\alpha -1) n+1}}}{r^3 (2 (\alpha -1) n+1)}\right)^{1/\alpha }}{\alpha +2 \alpha  n},
\end{equation}
\begin{equation}\label{5f}
p_t(r)=\frac{(\alpha -2) n \left(\frac{(2 n+1) r_0 \left(\frac{r}{r_0}\right)^{\frac{(2 \alpha -1) (2 n+1)}{2 (\alpha -1) n+1}}}{r^3 (2 (\alpha -1) n+1)}\right)^{1/\alpha }}{\alpha +2 \alpha  n}.
\end{equation}
The NEC corresponding to our anisotropic EoS model read as
\begin{equation}\label{5g}
\rho+p_r=\frac{2 (\alpha +\alpha  n-1) \left(\frac{(2 n+1) r_0 \left(\frac{r}{r_0}\right)^{\frac{(2 \alpha -1) (2 n+1)}{2 (\alpha -1) n+1}}}{r^3 (2 (\alpha -1) n+1)}\right)^{1/\alpha }}{\alpha +2 \alpha  n},
\end{equation}
\begin{equation}\label{5h}
\rho+p_t=\frac{(\alpha +(3 \alpha -2) n) \left(\frac{(2 n+1) r_0 \left(\frac{r}{r_0}\right)^{\frac{(2 \alpha -1) (2 n+1)}{2 (\alpha -1) n+1}}}{r^3 (2 (\alpha -1) n+1)}\right)^{1/\alpha }}{\alpha +2 \alpha  n}.
\end{equation}
At the throat, the above expressions reduce to the following,
\begin{equation}\label{5i}
\rho+p_r\mid_{r=r_0}=\frac{2 (\alpha +\alpha  n-1) \left(\frac{2 n+1}{r_0^2 (2 (\alpha -1) n+1)}\right)^{1/\alpha }}{\alpha +2 \alpha  n},
\end{equation}
\begin{equation}\label{5j}
\rho+p_t\mid_{r=r_0}=\frac{(\alpha +(3 \alpha -2) n) \left(\frac{2 n+1}{r_0^2 (2 (\alpha -1) n+1)}\right)^{1/\alpha }}{\alpha +2 \alpha  n}.
\end{equation}
We have presented the behavior of the energy density and radial NEC in Fig. \ref{fig:6} and \ref{fig:7}, respectively. We found that the energy density exhibits negative behavior corresponding to the GR case i.e. $\alpha=1$, whereas it is positive for other values of $\alpha$. Further, we found that NEC is violated in the neighborhood of the throat corresponding to every chosen value of $\alpha$ satisfying the constraints obtained in Table \ref{Table-1}. 

\begin{figure}[h]
\centering
\includegraphics[width=8cm, height=6cm]{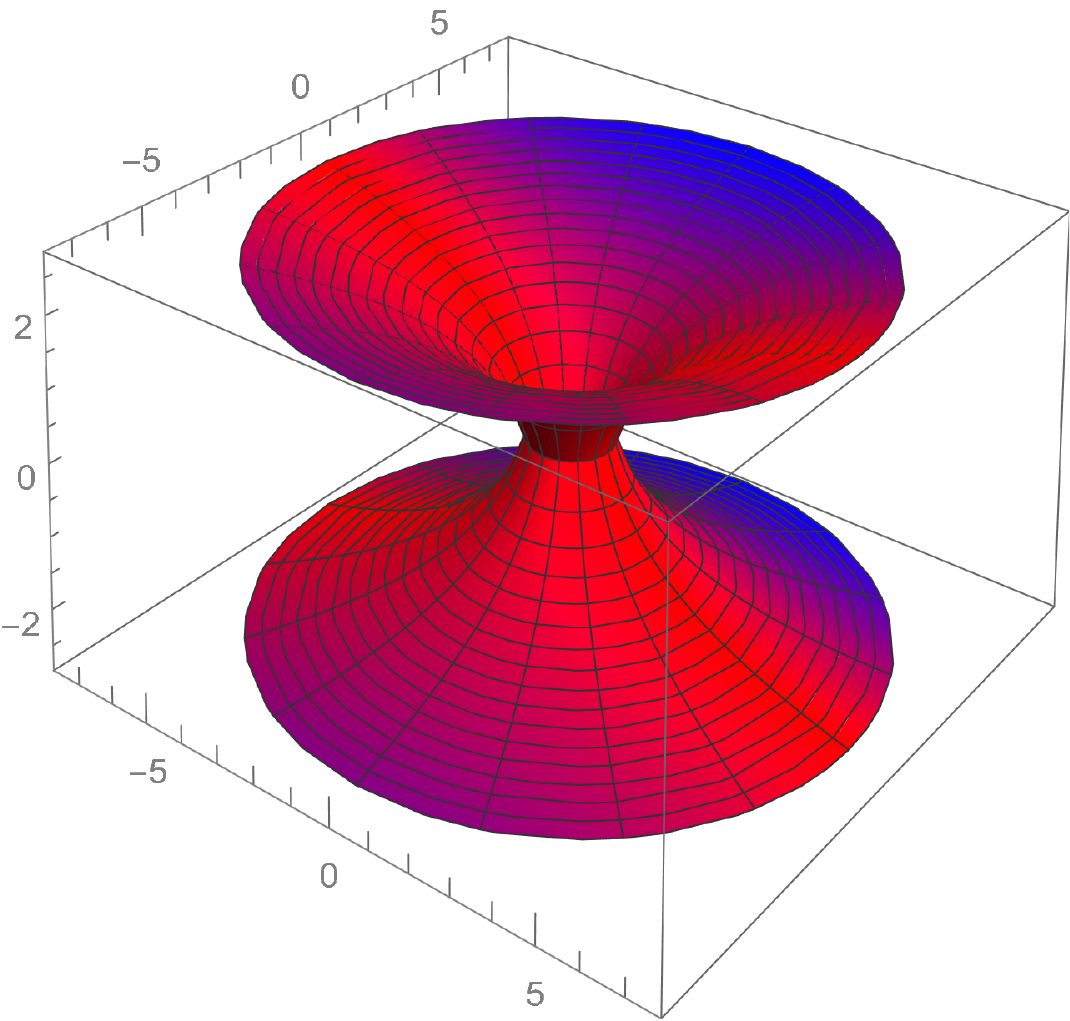}
\caption{Embedding diagram for the anisotropic EoS model corresponding to parameter values  $n=-1.5$ and $\alpha=0.85$.}
\label{fig:emb2}
\end{figure}

Further, we have investigated the 3D visualization of the surface sweep through a $2\pi$ rotation around the $z$-axis for this case, is shown in Fig. \ref{fig:emb2}.

\subsubsection{Isotropic EoS}\label{subsub13}
Usually, the energy-momentum tensor of the asymptotically flat wormholes in the GR is anisotropic i.e. $p_r\neq p_t$. 
We have already discussed the wormhole solutions for the anisotropic EoS case. In this section, we shall study wormhole solutions for the following isotropic relation 
\begin{equation}\label{44444}
    p_r=p_t.
\end{equation}
Now inserting the Eqs. \eqref{3i} and \eqref{3j} in the above expression \eqref{44444}, we could able to find the shape function
\begin{equation}
b(r)= c  \left( \frac{r}{c} \right)^3,
\end{equation}
where $c$ is the integrating constant. Imposing the throat condition $b(r_0)=r_0$ to the above expression, we obtain the final version of the shape function
\begin{equation}
b(r)= r_0  \left( \frac{r}{r_0} \right)^3.
\end{equation}
It is clear that the above expression is not asymptotically flat i.e. $\frac{b(r)}{r}\nrightarrow 0$ as $r\rightarrow \infty$. Thus it can be concluded that finding asymptotically flat wormhole solutions for isotropic pressure under constant redshift function for the $f(R,L_m)$ model \eqref{3g} is quite difficult.

\subsection{Model II}\label{subsec2}
We presume the following non-minimal type $f(R,L_m)$ function \cite{RV-2},
\begin{equation}
    f(R,L_m)=\frac{R}{2}+(1+\lambda\,R)L_m
\end{equation}
\justify where $\lambda$ is the coupling constant. The characteristics of Neutron stars have been investigated by using NICER data in the framework of the above-considered $f(R,L_m)$ model \cite{RV-2}. The coupling constant presents large values as compared to the weak-field limit. The same dependence appeared in scalar-tensor theories, where this parameter varies according to the scalar field mass, referred to as the chameleon mechanism \cite{KHR}. It is quite interesting to study the dependence of this coupling parameter in the context of different astrophysical systems such as white dwarfs, black holes, wormholes, and neutron stars. In particular, for the case $\lambda=0$, we retrieve the usual wormhole geometry of GR. The cosmological implications of the considered model have been tested in \cite{GM}. We will examine the characteristics of the given model in the context of wormholes.  \\
Therefore, we obtain the following field equations corresponding to this specific $f(R,L_m)$ function,  
\begin{widetext}
\begin{equation}\label{551}
\rho = \frac{\Phi ' \left(r \left(-b'\right)+2 r (r-b) \Phi '-3 b+4 r\right)+2 r (r-b) \Phi ''(r)+r^2 R}{2 \left(\lambda  \left(r \left(b'-4\right)+3 b\right) \Phi '+2 \lambda  r (b-r) \Phi '^2+r \left(2 \lambda  (b-r) \Phi ''+\lambda  r R+r\right)\right)},
\end{equation}
\begin{multline}\label{552}
P_r=\frac{1}{\mathcal{K}}\left[r \left(b' \left((\lambda  r R+r) \Phi '+4 \lambda  R+2\right)-r \left(2 (\lambda  r R+r) \Phi '^2-4 \lambda  R \Phi '(r)+r (\lambda  R+1) \left(2 \Phi ''+R\right)\right)\right)\right.\\\left.
+b \left(r \left(\Phi ' \left(2 (\lambda  r R+r) \Phi '-5 \lambda  R-1\right)+2 r (\lambda  R+1) \Phi ''\right)-4 \lambda  R-2\right)\right],
\end{multline}
\begin{multline}\label{553}
P_t=\frac{1}{\mathcal{K}}\left[r \left(b' \left(-\lambda  r R \Phi '+2 \lambda  R+1\right)+2 \lambda  r^2 R \left(\Phi '^2+\Phi ''\right)-r \left(r R (\lambda  R+1)+2 \Phi '\right)\right)\right.\\\left.
+b \left(r \left(\Phi ' \left(-2 \lambda  r R \Phi '+\lambda  R+2\right)-2 \lambda  r R \Phi ''\right)+2 \lambda  R+1\right)\right],
\end{multline}
\end{widetext}
where
\begin{multline}
\mathcal{K}=2 r (\lambda  R+1) \left(\lambda  \left(r \left(b'-4\right)+3 b\right) \Phi ' \right.\\\left.
+2 \lambda  r (b-r) \Phi '^2+r \left(2 \lambda  (b-r) \Phi ''+\lambda  r R+r\right)\right).
\end{multline}

\subsubsection{ $b(r)=r_0+\gamma ^2 r_0 \left(1-\frac{r_0}{r}\right)$}\label{subsub21}
\begin{figure}[h]
    \centering
    \includegraphics[scale=0.59]{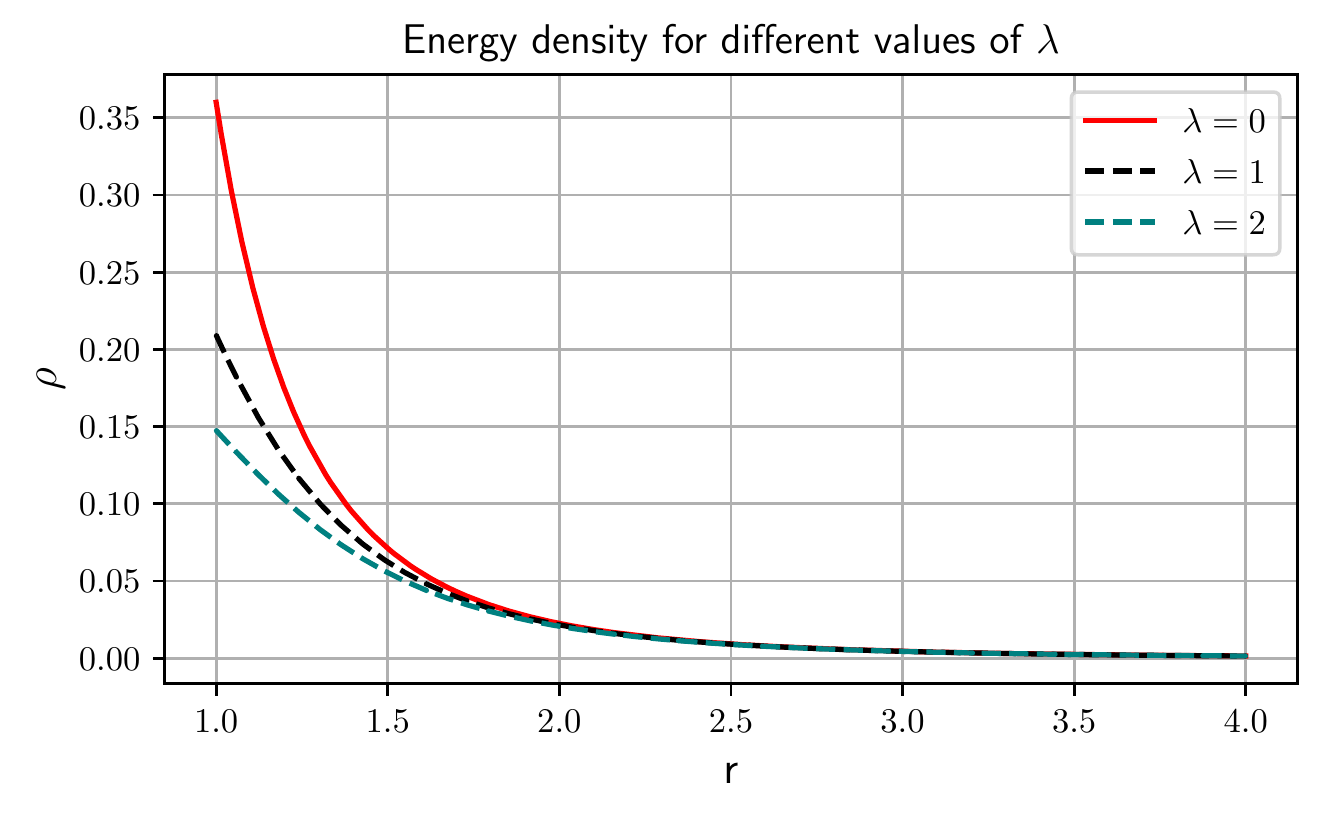}
    \caption{Energy density with $r_0=1$ and $\gamma=0.6$. Note that for $\lambda=0$ corresponds to GR case.}
    \label{fig:5a}
\end{figure}

\begin{figure}[h]
    \centering
    \includegraphics[scale=0.59]{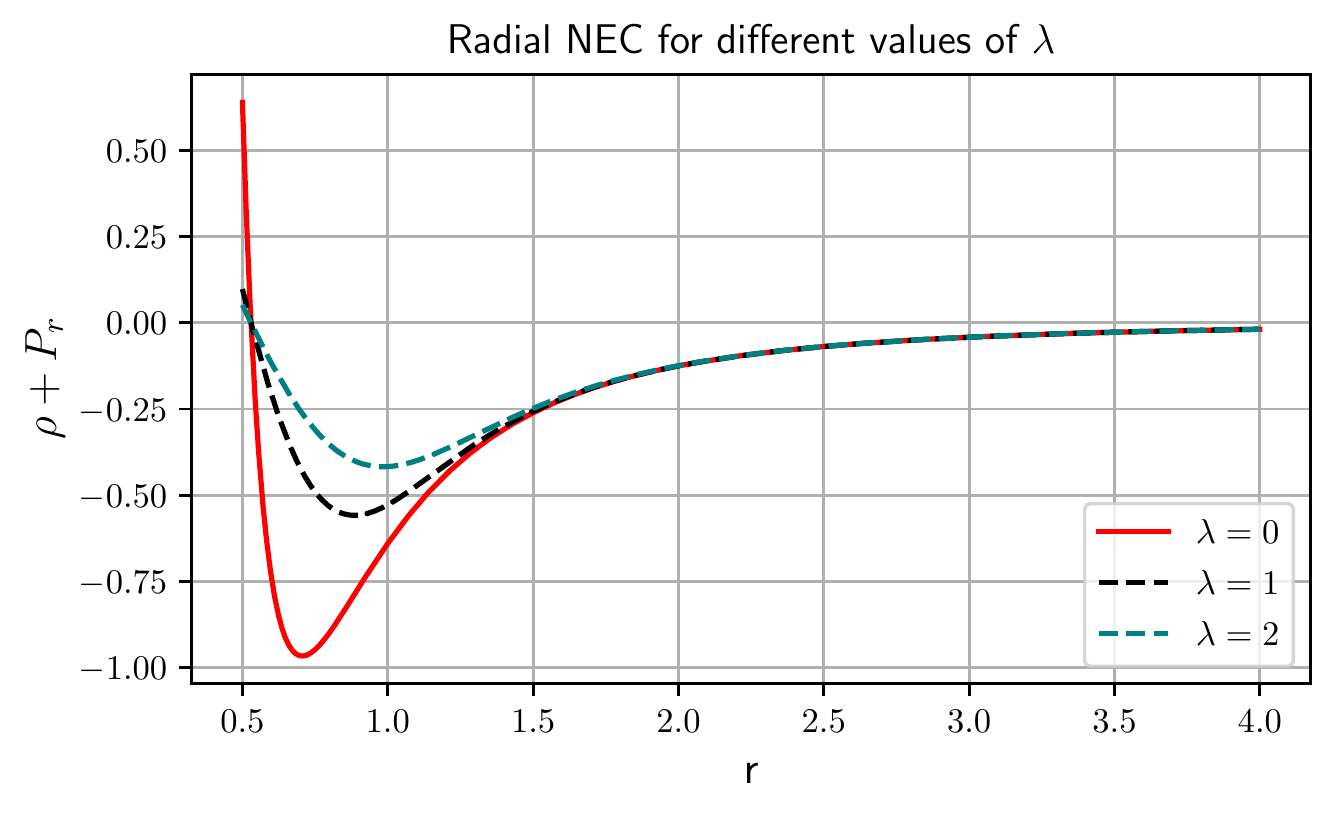}
    \caption{Radial NEC with $r_0=1$ and $\gamma=0.6$. Note that for $\lambda=0$ corresponds to GR case.}
    \label{fig:5aa}
\end{figure}

Here, we turn our attention to the model with $b(r)=r_0+\gamma ^2 r_0 \left(1-\frac{r_0}{r}\right)$ \cite{Lobo2009}, where $0<\gamma<1$ is particularly interesting to have wormhole solutions that satisfy the flaring-out condition at wormhole throat. The stress-energy tensor profile for this specific function under constant redshift function is given by
\begin{equation}
\rho=\frac{\gamma ^2 r_0^2}{r^4+2 \gamma ^2 \lambda  r_0^2},
\end{equation}
\begin{equation}
P_r=\frac{1}{\mathcal{L}^2}\left[2 \gamma ^2 \lambda  r_0^3 \left(3 \gamma ^2 r_0-2 \left(\gamma ^2+1\right) r\right)-r^4 r_0 \mathcal{M}\right],
\end{equation}
\begin{equation}
P_t=\frac{r^4 r_0 \left(\gamma ^2 (r-2 r_0)+r\right)+4 \gamma ^2 \lambda  r_0^3 \mathcal{M}}{2 \mathcal{L}^2},
\end{equation}
where $\mathcal{L}=\left(r^4+2 \gamma ^2 \lambda  r_0^2\right)$ and $\mathcal{M}=\left(\gamma ^2 (r-r_0)+r\right)$.\\
For the case of the NEC along the radial and tangential direction is provided by
\begin{equation}\label{5gg}
\rho+P_r=-\frac{r_0 \left(\gamma ^2 (r-2 r_0)+r\right) \left(r^4+4 \gamma ^2 \lambda  r_0^2\right)}{\mathcal{L}^2},
\end{equation}
\begin{equation}
\rho+P_t=\frac{\left(\gamma ^2+1\right) r r_0 \left(r^4+4 \gamma ^2 \lambda  r_0^2\right)}{2 \mathcal{L}^2}.
\end{equation}
One can find that at wormhole throat, the above equations reduce to
\begin{equation}\label{5a1}
\rho+P_r\mid_{r=r_0}=-\frac{\left(1-\gamma ^2\right) \left(4 \gamma ^2 \lambda +r_0^2\right)}{\left(2 \gamma ^2 \lambda +r_0^2\right)^2},
\end{equation}
\begin{equation}\label{5a2}
\rho+P_t\mid_{r=r_0}=\frac{\left(\gamma ^2+1\right) \left(4 \gamma ^2 \lambda +r_0^2\right)}{2 \left(2 \gamma ^2 \lambda +r_0^2\right)^2}.
\end{equation}
For this specific case, we noticed that the RHS of Eq. \eqref{5a1} is a negative quantity for any $\lambda >0$ and $0<\gamma<1$, which confirms the violation of radial NEC at the throat. However, from Eq. \eqref{5a2}, we can see that tangential NEC is satisfied at the throat. The graphical representations of $\rho$ and $\rho+P_r$ are shown in Fig. \ref{fig:5a} and \ref{fig:5aa}. Interestingly, energy density is positive throughout space-time, whereas NEC is violated at the throat and its neighborhood. Also, we observed that for substantial values of $\lambda$, the radial NEC $\rho+P_r$ would be validated.

\subsubsection{$b(r)=r\,e^{r_0-r}$}\label{subsub22}
\begin{figure}[h]
    \centering
    \includegraphics[scale=0.59]{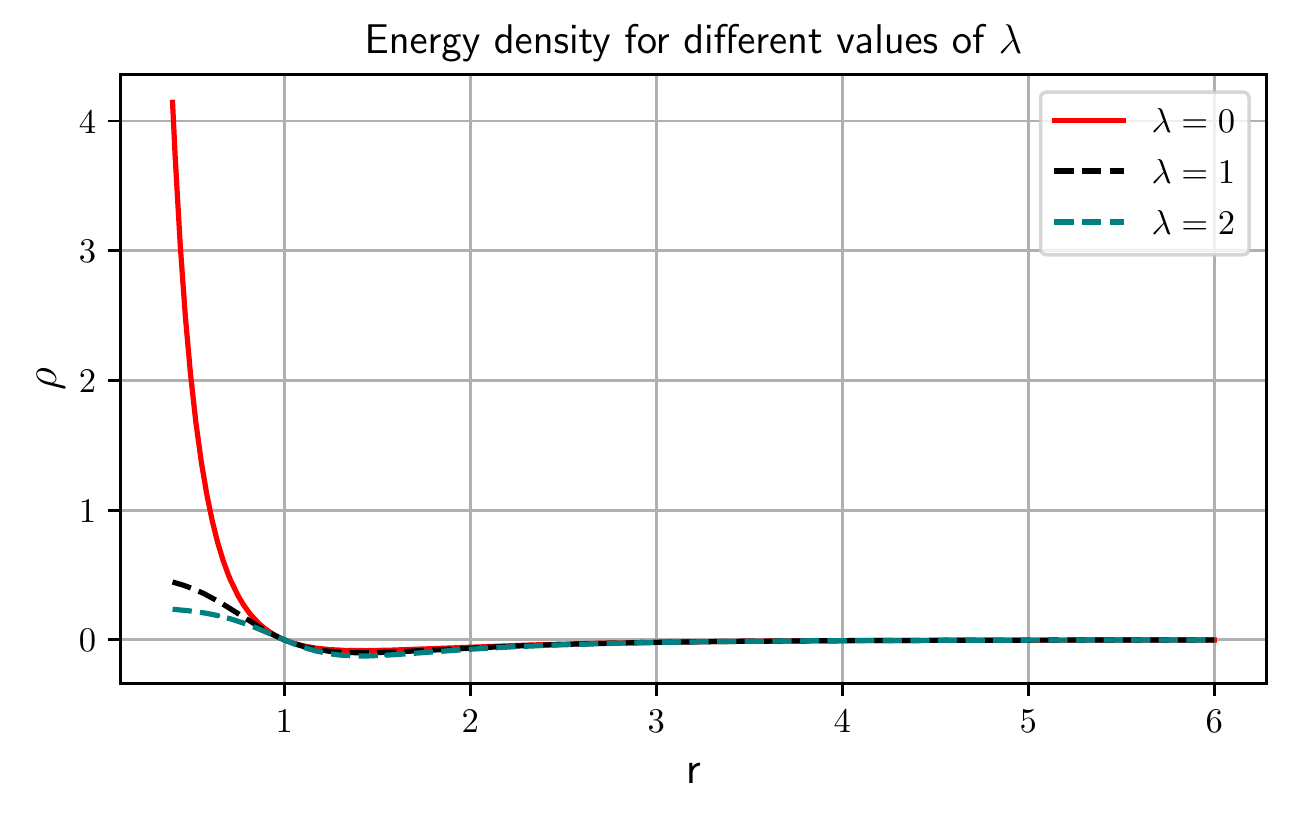}
    \caption{Energy density with $r_0=0.5$. Note that for $\lambda=0$ corresponds to GR case.}
    \label{fig:5b}
\end{figure}

\begin{figure}[h]
    \centering
    \includegraphics[scale=0.59]{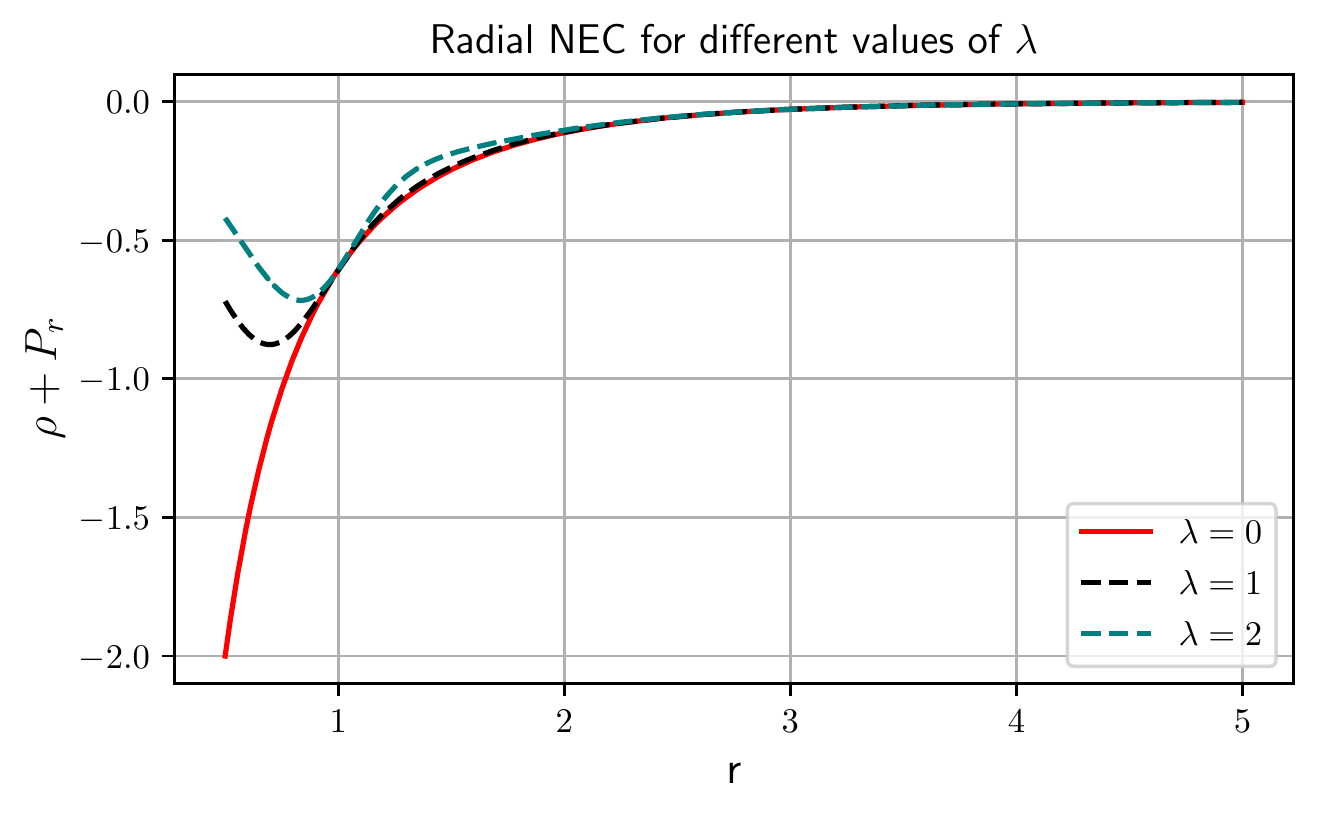}
    \caption{Radial NEC with $r_0=0.5$. Note that for $\lambda=0$ corresponds to GR case.}
    \label{fig:5bb}
\end{figure}

Considering the specific choice for the form function $b(r)=r\,e^{r_0-r}$ \cite{Bamba1} where $0<r_0<1$ is particularly interesting to have wormhole solutions that satisfy the condition $b^{'}(r_0)<1$. Taking this form into account, the generalized field equations Eqs. (\ref{551}-\ref{553}) can be read as
\begin{equation}\label{5b1}
\rho=\frac{(r-1) e^{r_0}}{-e^r r^2+2 \lambda  r e^{r_0}-2 \lambda  e^{r_0}},
\end{equation}
\begin{equation}\label{5b2}
P_r=\frac{e^{r_0} \left(2 \lambda  \left(r^2-1\right) e^{r_0}-e^r r^2\right)}{\left(e^r r^2-2 \lambda  (r-1) e^{r_0}\right)^2},
\end{equation}
\begin{equation}\label{5b3}
P_t=\frac{e^{r_0} \left(e^r r^3-4 \lambda  (r-1) e^{r_0}\right)}{2 \left(e^r r^2-2 \lambda  (r-1) e^{r_0}\right)^2}.
\end{equation}
Using Eqs. (\ref{5b1}-\ref{5b3}), one can obtain the NEC given by
\begin{equation}\label{5b4}
\rho+P_r=\frac{r e^{r_0} \left(4 \lambda  (r-1) e^{r_0}-e^r r^2\right)}{\left(e^r r^2-2 \lambda  (r-1) e^{r_0}\right)^2},
\end{equation}
\begin{equation}\label{5b5}
\rho+P_t=\frac{(r-2) e^{r_0} \left(4 \lambda  (r-1) e^{r_0}-e^r r^2\right)}{2 \left(e^r r^2-2 \lambda  (r-1) e^{r_0}\right)^2}.
\end{equation}
For this wormhole model, we assume the throat radius $r_0=0.5$ and hence the flare-out condition $b^{'}(0.5)=0.5<1$ obeyed at the throat. Also, we can see from Eq. \eqref{5b4} that
$\rho+P_r\mid_{r=r_0}=-\frac{r_0 \left(4 \lambda +r_0^2-4 \lambda  r_0\right)}{\left(2 \lambda +r_0^2-2 \lambda  r_0\right)^2},$
which can be violated for $0<r_0<1$ and $\lambda >\frac{r_0^2}{4 (r_0-1)}$. Also, we noticed that radial NEC could be validated at the throat for $0<r_0<1$ and $\lambda <\frac{r_0^2}{2( r_0-1)}\lor \frac{r_0^2}{2 (r_0-1)}<\lambda <\frac{r_0^2}{4 (r_0-1)}$. Further, we study tangential NEC at $r=r_0$ provided by $\rho+P_t=-\frac{(r_0-2) \left(4 \lambda +r_0^2-4 \lambda  r_0\right)}{2 \left(2 \lambda +r_0^2-2 \lambda  r_0\right)^2}$. It is observed that tangential NEC is violated for $0<r_0<1$ within the range $\lambda <\frac{s^2}{2 s-2}\lor \frac{r_0^2}{2 r_0-2}<\lambda <\frac{r_0^2}{4 r_0-4}$, whereas satisfied within $\lambda >\frac{r_0^2}{4 (r_0-1)}$.
We have presented the graphical behavior of energy density and NEC in Figs. \ref{fig:5b} and \ref{fig:5bb} with some particular values discussed above. It is obvious that NEC is violated, whereas energy density is respected in the vicinity of the throat. However, one can notice that energy density is no longer validated far from the throat.


% \section{Embedding diagram}\label{sec5}
% In this section, we will construct the embedding diagrams corresponding to obtained wormhole solution for both models. We consider the equatorial slice $\theta=\pi/2$ for a fixed time i.e. $t=$constant. Hence, equation \eqref{3a} reduces to
% \begin{equation}
% \label{6a}
% ds^2=\left(1-\frac{b(r)}{r}\right)^{-1}dr^2+r^2 d\phi^2.
% \end{equation}
% The above metric can be embedded into three-dimensional Euclidean space with cylindrical coordinates $r,\,\ \phi$ and $z$ as
% \begin{equation}
% \label{6b}
% ds^2=dz^2+dr^2+r^2 d\phi^2.
% \end{equation}
% Now, on comparing equations \eqref{6a} and \eqref{6b}, we obtained the following slope equation so that by integrating it, one can find the embedding surface $z(r)$,
% \begin{equation}
% \label{6c}
% \frac{dz}{dr}=\pm \sqrt{\frac{r}{r-b(r)}-1}.
% \end{equation}
% We have presented the embedding diagrams in Figs. \ref{fig:emb1} and \ref{fig:emb2} for the wormhole solutions obtained in equations \eqref{4c} and \eqref{5c} respectively.

\section{Amount of exotic matter}\label{sec6}
\begin{figure}[h]
    \centering
    \includegraphics[scale=0.6]{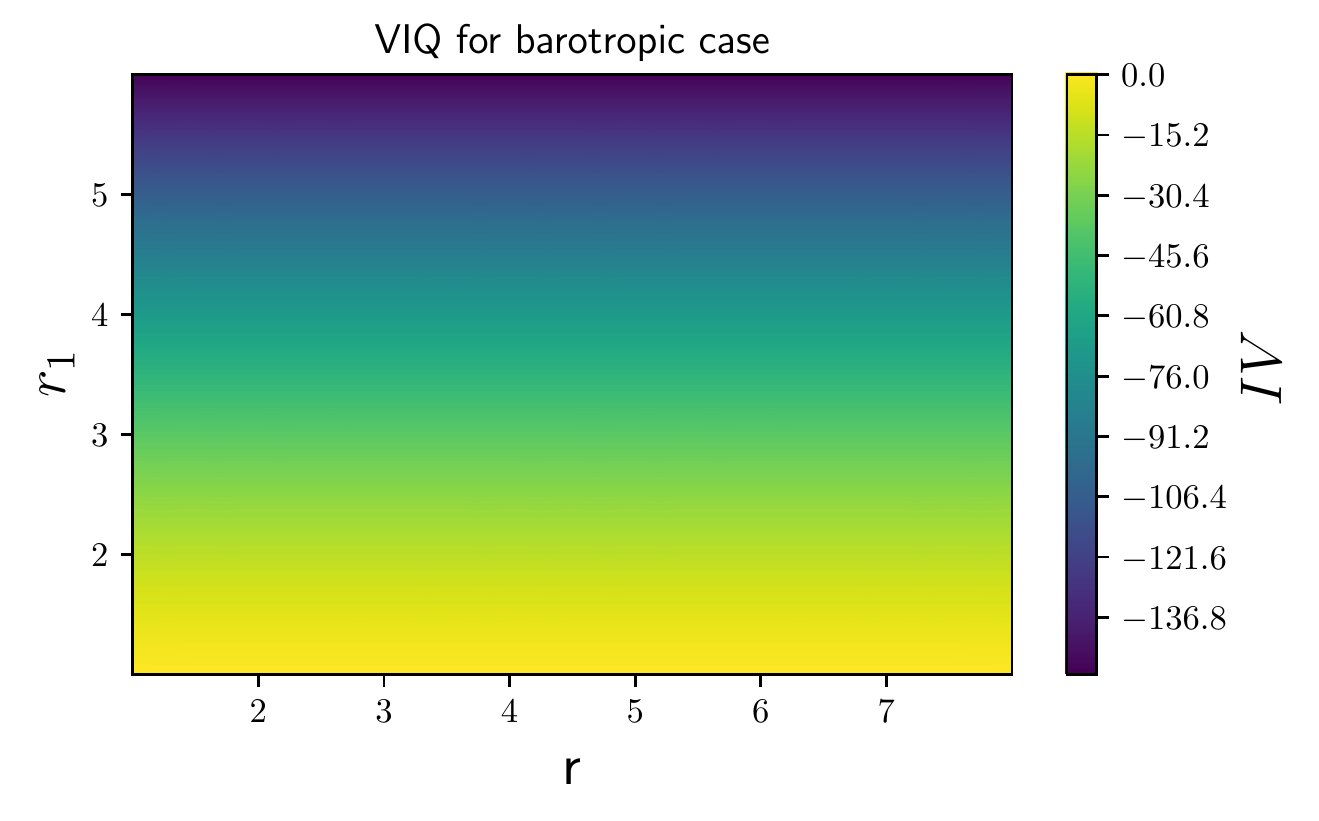}
    \caption{Profile of VIQ with $\alpha=2$, $\omega=-2$, and $r_0=1$.}
    \label{fig:10}
\end{figure}

\begin{figure}[h]
    \centering
    \includegraphics[scale=0.6]{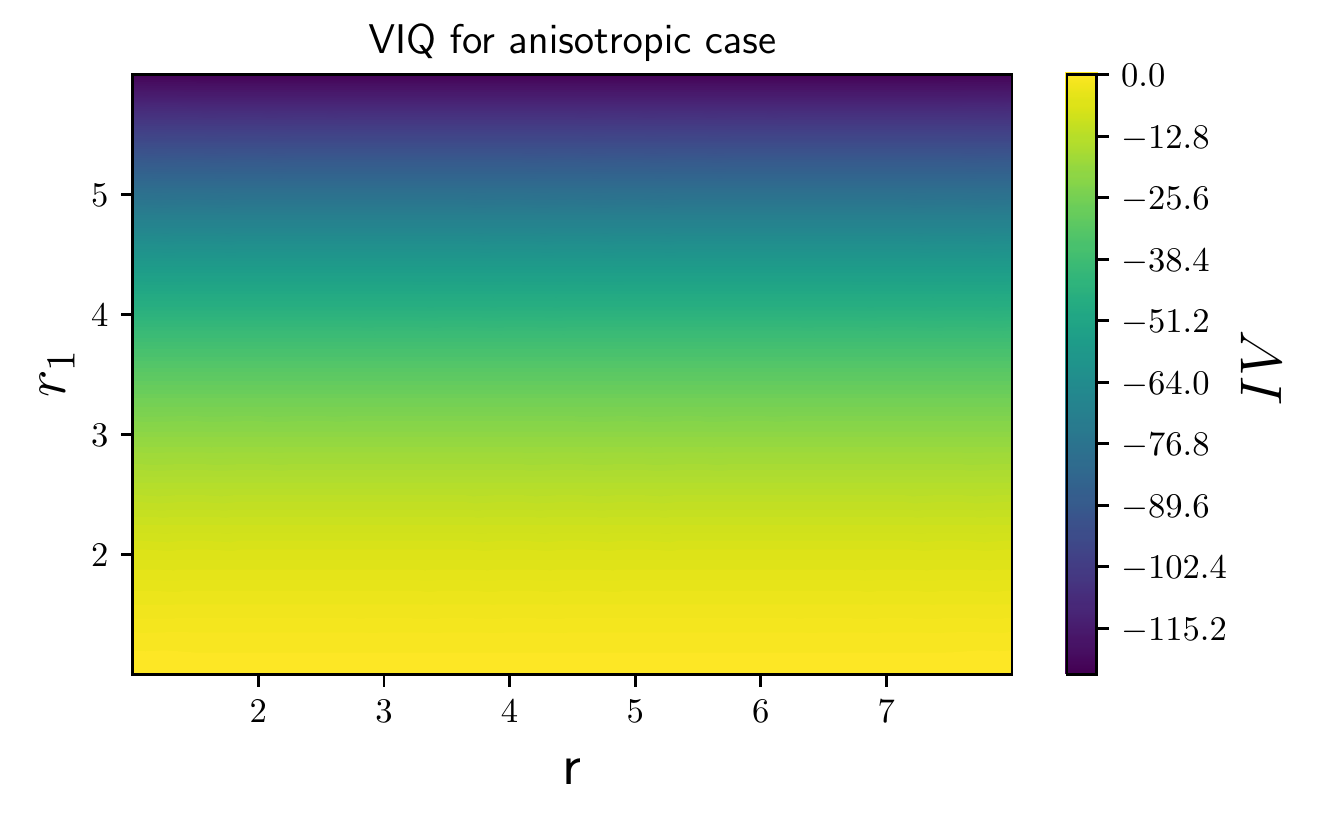}
    \caption{Profile of VIQ with $\alpha=4$, $n=-0.7$ and $r_0=1$.}
    \label{fig:11}
\end{figure}

\begin{figure}[h]
    \centering
    \includegraphics[scale=0.6]{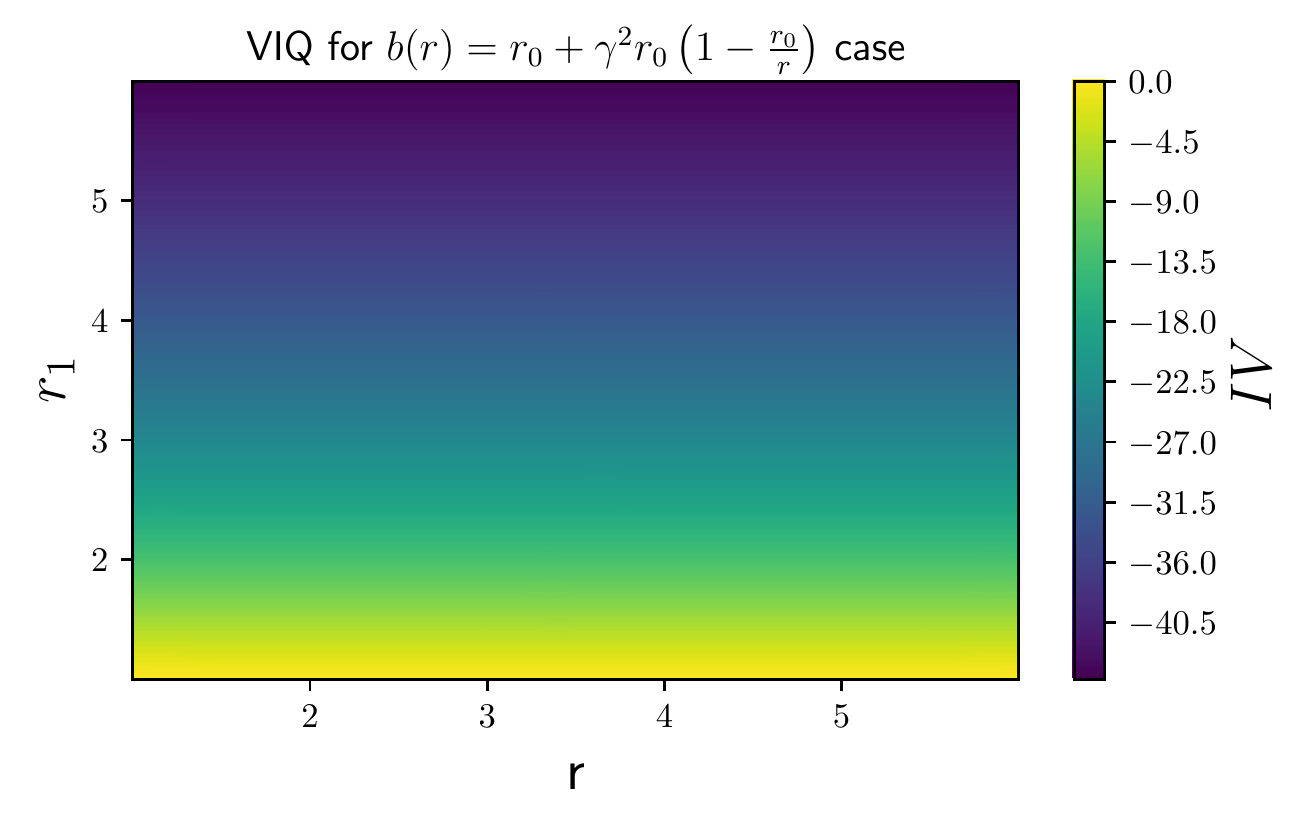}
    \caption{Profile of VIQ with $\gamma=0.6$, $\lambda=2$ and $r_0=1$.}
    \label{fig:111}
\end{figure}

Now we will estimate the amount of exotic matter required for a wormhole to be stable. To do this, we use the volume integral quantifier (VIQ) approach, introduced by Visser et al. \cite{Visser3}, which can quantify the average amount of matter present in spacetime, violating NEC. The VIQ is defined as 
\begin{equation}
IV=\oint [\rho+P_r]dV
%=2\int_{r_0}^{\infty}(\rho+P_r)dV,
\end{equation}
where the volume can be read as $dV=r^2\,dr\,d\Omega$ with $d\Omega$ the solid angle. Since $\oint dV=2\int_{r_0}^{\infty}dV=8\pi \int_{r_0}^{\infty}r^2dr,$ we have
%which can also be written as
\begin{equation}
IV=8\pi \int_{r_0}^{\infty}(\rho+P_r)r^2dr.
\end{equation}
Now, the volume integral corresponding to a wormhole whose field varies from the throat $r_0$ to a fixed radius $r_1$ with $r_1\geq r_0,$ is given as
\begin{equation}\label{1122}
IV=8\pi \int_{r_0}^{r_1}(\rho+P_r)r^2dr.
\end{equation}
Now with the help of the above Eq. \eqref{1122}, we have investigated the volume integral and presented the nature of those in Figs. (\ref{fig:10}-\ref{fig:111}). It is evident from the graphs that $IV\rightarrow 0$, as $r_1\rightarrow r_0$. Thus, we can conclude that a small fraction of exotic matter can stabilize a traversable wormhole. We found that one can minimize the total amount of average null energy condition (ANEC) violating matter by taking suitable wormhole geometry. One can check the references \cite{Baransky,Channuie1} to review some interesting applications of VIQ.

\section{Conclusions}\label{sec7}
A wormhole is a short-cut that act as a subway for distinct regions of spacetime or distinct spacetimes apart from each other. The theoretical solution of field equations of GR that violates energy conditions gave rise to the concept of a wormhole. Nowadays, wormhole solutions in the context of modified gravity theories have attracted the attention of researchers. Modified gravity theories play a significant role in bypassing the necessity of undetected exotic matter in constructing wormholes. In this paper, we have examined different wormhole solutions in the $f(R,L_m)$ gravity. This theory has been investigated to analyze some dark energy candidates by utilizing dynamical system technique \cite{Azevedo}. Further, hydrostatic equilibrium configurations of neutron stars in the context of $f(R,L_m)$ gravity is obtained in \cite{Carvalho}. Moreover, $f(R,L_m)$ gravity describes pulsars as massive as PSR J2215+5135 \cite{Linares} with the help of an equation of state for nuclear matter. These attractive features of $f(R,L_m)$ gravity motivated us to study wormholes in this gravity.\\
In this manuscript, we attempted to explore wormhole geometry in the framework of $f(R,L_m)$ gravity. We have derived the field equations for the wormhole metric \eqref{3a} in the $f(R,L_m)$ gravity background. Then we have considered two different $f(R,L_m)$ models, such as, $f(R,L_m)=\frac{R}{2}+L_m^\alpha$ (model I) and $ f(R,L_m)=\frac{R}{2}+(1+\lambda\,R)L_m$ (model II), where $\alpha$ and $\lambda$ are arbitrary model parameters. Firstly, we have studied the wormhole solutions for three different cases (i) a linear barotropic EoS, (ii) an anisotropic EoS, and (iii) isotropic EoS, under model I.
For the linear barotropic case, we calculated the shape function $b(r)$ and obtained the possible regions where the flatness condition is satisfied. We found the flaring-out condition is satisfied for any $\omega<-1$ corresponding to $\alpha>-\frac{1}{\omega-1}$ under the asymptotic background. Further, we have investigated the energy conditions and noticed that energy density is positive in the entire space-time. NEC is violated in the neighborhood of the throat for increasing values of $\alpha$ under the phantom region. However, the violation of NEC may not be maintained for very large values of $\alpha$.\\
Again, for the anisotropic case, we have obtained the possible ranges of $\alpha$ and $n$ using the asymptotically flatness condition, which is listed in Table \ref{Table-1}. By setting the appropriate values from the Table \ref{Table-1}, we have studied the behavior of the shape functions and energy conditions. We found that the flaring-out condition is satisfied at the wormhole throat. The energy density exhibit negative behavior corresponding to the GR case, i.e., $\alpha=1$, while it is positive for other values of $\alpha$. The NEC is violated in the neighborhood of the throat corresponding to every chosen value of $\alpha$ satisfying the constraints obtained in Table \ref{Table-1}. Further, we have investigated wormholes for the isotropic relation and observed that the obtained shape function is not asymptotically flat. It seems that finding asymptotically flat wormhole solutions for isotropic pressure under constant redshift function in $f(R,L_m)$ scenario is quite difficult.\\
Further, finding the wormhole solutions for model II in an analytic way is quite tricky due to the complexity of the field equations. Therefore, we have studied wormhole solutions with two different shape functions, i.e., $b(r)=r_0+\gamma ^2 r_0 \left(1-\frac{r_0}{r}\right)$ and $b(r)=r\,e^{r_0-r}$ in order to investigate model II. For $b(r)=r_0+\gamma ^2 r_0 \left(1-\frac{r_0}{r}\right)$ case, we found that radial NEC is violated, and tangential NEC is satisfied near the throat within $\lambda>0$ and $0<\gamma<1$. Energy density is positive in the entire spacetime. Again, for $b(r)=r\,e^{r_0-r}$, we investigated the NEC thoroughly and obtained some validity ranges for NEC. It is observed radial NEC is violated within $0<r_0<1$ and $\lambda >\frac{r_0^2}{4 (r_0-1)}$ whereas, tangential NEC is violated for $0<r_0<1$ within the range $\lambda <\frac{r_0^2}{2 r_0-2}\lor \frac{r_0^2}{2 r_0-2}<\lambda <\frac{r_0^2}{4 r_0-4}$. In this case, we found that energy density is positive in the neighborhood of the throat. Moreover, we noticed that the value of the coupling constant obtained in our study is consistent with the values obtained in \cite{RV-1,RV-2}. According to the Ref. \cite{RV-2}, the best result can be achieved when the coupling parameter $\lambda$ is around $30$. Further, in case of white dwarf system the consistent value of $\lambda$ lies between $0$ and $1$ \cite{RV-1}. However, in our study, we noticed that for a traversable wormhole solution, $\lambda$ is a non-negative value and that is consistent with values obtained in the above results.\\
In addition, we have examined the VIQ to study the amount of exotic matter required at the throat for a traversable wormhole. In our analysis, we found that a small amount of exotic matter is necessary for a traversable wormhole. We can conclude that the modification of standard GR can efficiently minimize the use of exotic matter and provide a stable traversable wormhole solution. Although wormholes have not been detected yet, in this study, we have investigated the possible existence of wormhole geometries in the context of $f(R,L_m)$ gravity. In this study, we have considered the constant redshift function, i.e., $\Phi(r)=\text{constant}$. It would be interesting to investigate wormholes with non-constant redshift functions in this modified gravity in the near future.

\section*{Data Availability Statement}
There are no new data associated with this article.

\section*{Acknowledgments}

RS acknowledges University Grants Commission (UGC), New Delhi, India, for awarding a Senior Research Fellowship (UGC-Ref. No.: 191620096030). ZH acknowledges the Department of Science and Technology (DST), Government of India, New Delhi, for awarding a Senior Research Fellowship (File No. DST/INSPIRE Fellowship/2019/IF190911). P.K.S. acknowledges the National Board for Higher Mathematics (NBHM) under the Department of Atomic Energy (DAE), Govt. of India for financial support to carry out the Research project No.: 02011/3/2022 NBHM(R.P.)/R\&D II/2152 Dt.14.02.2022 and Transilvania University of Brasov for Transilvania Fellowship for Visiting Professors. We are very much grateful to the honorable referees and to the editor for the illuminating suggestions that have significantly improved our work in terms
of research quality, and presentation.


\begin{thebibliography}{52}
\footnotesize
\bibitem{Flamm} L. Flamm, \textit{Phys. Z.} \textbf{17}, 448 (1916).
\bibitem{Einstein} A. Einstein and N. Rosen, \textit{Phys. Rev.} \textbf{48}, 73 (1935).
\bibitem{Fuller} R. W. Fuller, J. A. Wheeler, \textit{Phys. Rev.} \textbf{128}, 919 (1962).
\bibitem{Morris} M. S. Morris, K. S. Thorne, U. Yurtsever, \textit{Phys. Rev. Lett.} \textbf{61}, 1446 (1988).
\bibitem{Thorne} M. S. Morris, K. S. Thorne, \textit{Am. J. Phys.} \textbf{56}, 395 (1988).
\bibitem{Visser} M. Visser, \textit{Lorentzian wormholes: from Einstein to Hawking} (AIP Press, New York, 1995).
\bibitem{Wheeler} J. A. Wheeler, \textit{Geons. Phys. Rev.} \textbf{97}, 511 (1955).
\bibitem{Bronnikov1} K. A. Bronnikov, S. V. Grinyok, \textit{Gravit. Cosmol.} \textbf{7}, 297 (2001).
\bibitem{Bronnikov2} K. A. Bronnikov, A. A. Starobinsky, \textit{JETP Lett.} \textbf{85}, 1 (2007).
\bibitem{Bronnikov3} J.A. Gonzalez, F.S. Guzman,O. Sarbach, \textit{Class. Quantum Gravity} \textbf{26}, 015010 (2009).
\bibitem{Oliveira} F. S. N. Lobo, M.A. Oliveira, \textit{Phys. Rev. D} \textbf{80}, 104012 (2009).
\bibitem{Halilsoy} S.H. Mazharimousavi, M. Halilsoy, \textit{Mod. Phys. Lett. A} \textbf{31}, 1650192 (2016).
\bibitem{Azizi} T. Azizi, \textit{Int. J. Theor. Phys.} \textbf{52}, 3486 (2013).
\bibitem{Kim} K. A. Bronnikov, S.W. Kim, \textit{Phys. Rev. D} \textbf{67}, 064027 (2003).

\bibitem{Camera} M. L. Camera, \textit{Phys. Lett. B} \textbf{573}, 27 (2003).
\bibitem{Riazi} F. Parsaei, N. Riazi, \textit{Phys. Rev. D} \textbf{91}, 024015 (2015).
%\bibitem{Gupta} S. Kar, S. Lahir, S. S. Gupta, \textit{Phys. Lett. B} \textbf{750}, 319 (2016).
\bibitem{Garcia1} N. M. Garcia, F. S. N. Lobo, \textit{Phys. Rev. D} \textbf{82}, 104018 (2010).
\bibitem{Garcia2} N. M. Garcia, F. S. N. Lobo, \textit{Class. Quantum Gravity} \textbf{28}, 085018 (2011).
\bibitem{O.B.}  O. Bertolami et al., \textit{Phys. Rev. D} \textbf{75}, 104016 (2007). 
\bibitem{THK} T. Harko, \textit{Phys. Lett. B} \textbf{669}, 376 (2008).
\bibitem{THK-2} T. Harko, \textit{Phys. Rev. D} \textbf{81}, 084050 (2010).
\bibitem{THK-3} T. Harko, \textit{Phys. Rev. D} \textbf{81}, 044021 (2010).
\bibitem{THK-4} T. Harko, \textit{Phys. Rev. D} \textbf{90}, 044067 (2014).
\bibitem{THK-5} T. Harko and S. Shahidi, \textit{Eur. Phys. J. C} \textbf{82}, 219 (2022).
\bibitem{V.F.-2} V. Faraoni, \textit{Phys. Rev. D} \textbf{76}, 127501 (2007).
\bibitem{THK-6} T. Harko and F. S. N. Lobo, \textit{Eur. Phys. J. C} \textbf{70}, 373-379 (2010).
\bibitem{FR} V. Faraoni, \textit{Cosmology in Scalar-Tensor Gravity}, Kluwer Academic, Dordrecht (2004).
\bibitem{JP}  O. Bertolami., J. Páramos, and S. Turyshev, \textit{arXiv}, arXiv:gr-qc/0602016.
\bibitem{GM} B.S. Goncalves and P.H.R.S. Moraes, \textit{arXiv}, arXiv:2101.05918.
\bibitem{RV-1} R. V. Labato et al., \textit{Eur. Phys. J. C} \textbf{82}, 540 (2022).
\bibitem{RV-2} R. V. Labato, G. A. Carvalho, and C. A. Bertulani, \textit{Eur. Phys. J. C} \textbf{81}, 1013 (2021).
\bibitem{THK-7} T. Harko and S. Shahidi, \textit{Eur. Phys. J. C} \textbf{82}, 1003 (2022).
\bibitem{THK-8} T. Harko and M. J. Lake, \textit{Eur. Phys. J. C} \textbf{75}, 60 (2015).

\bibitem{Gonzalez} J.A. Gonzalez, F. S. Guzman, N. Montelongo-Garcia, and T. Zannias, \textit{Phys. Rev. D} \textbf{79}, 064027 (2009).
\bibitem{Lobooo} F. S. N. Lobo, \textit{Phys. Rev. D} \textbf{71}, 084011 (2005).
\bibitem{LB}  T. Harko and F. S. N. Lobo, \textit{Galaxies} \textbf{2014(2)}, 410-465 (2014).
\bibitem{Jay-2} L. V. Jaybhaye et al., \textit{Phys. Lett. B} \textbf{831}, 137148 (2022).
\bibitem{Sube} A. Bose, G. Sardar, and S. Chakraborty, \textit{Phys. Dark Univ.} \textbf{37}, 101087 (2022).
\bibitem{Lobo1} F. S. N. Lobo, \textit{Phys. Rev. D} \textbf{71}, 084011 (2005).
\bibitem{Lobo2} F. S. N. Lobo, F. Parsaei and N. Riazi, \textit{Phys. Rev. D} \textbf{87}, 084030 (2013).
\bibitem{Mandal} Z. Hassan, S. Mandal and P. K. Sahoo, \textit{Forts. Phys.} \textbf{69}, 2100023 (2021).
\bibitem{Jasim} Ksh. N. Singh, A. Banerjee, F. Rahaman, and M. K. Jasim  \textit{Phys. Rev. D} \textbf{101}, 084012 (2020).
%\bibitem{Sarker}  F. Rahaman, M. Kalam, M. Sarker, A. Ghosh, and B. Raychaudhuri, \textit{, Gen. Relativ. Gravit.} \textbf{39}, 145 (2007).
\bibitem{Moraes1} P. H. R. S. Moraes and P. K. Sahoo, \textit{Phys. Rev. D} \textbf{96}, 044038 (2017).
\bibitem{KHR} J. Khoury and A. weltman, \textit{Phys. Rev. Lett.} \textbf{93}, 171104 (2004).
\bibitem{Lobo2009} F. S. N. Lobo and M. A. Oliveira, \textit{Phys. Rev. D} \textbf{80}, 104012 (2009).
\bibitem{Bamba1} G.C. Samanta, N. Godani, K. Bamba, \textit{Int. J. Mod. Phys. D} \textbf{29}, 2050068 (2020)
\bibitem{Visser3} M. Visser, S. Kar and N. Dadhich, \textit{Phys. Rev. Lett.} \textbf{90}, 201102 (2003).
\bibitem{Channuie1} K. Jusufi, P. Channuie, M. Jamil, \textit{Eur. Phys. J. C} \textbf{80}, 127 (2020).
%\bibitem{adm2} P. T. Chrusciel, Lectures on Energy in General Relativity, Krakow, 2010.
\bibitem{Baransky} O. Sokoliuk, S. Mandal, P. K. Sahoo, A. Baransky, \textit{Eur. Phys. J. C} \textbf{82}, 280 (2022).
\bibitem{Azevedo}  R. P. L. Azevedo and J. Paramos, \textit{Phys. Rev. D} \textbf{94}, 064036 (2016).
\bibitem{Carvalho}  G. A. Carvalho et al., \textit{Eur. Phys. J. C} \textbf{80}, 483 (2020).
\bibitem{Linares} M. Linares et al. \textit{Astrophys. J.} \textbf{859}, 54 (2018).

\end{thebibliography}
\end{document}